\documentclass[12pt]{iopart}
\usepackage{iopams}
\usepackage{amsopn}
\usepackage{setstack}
\usepackage{bbm}
\usepackage[german,english]{babel}  
\usepackage[T1]{fontenc}
\usepackage{url} 
\usepackage{dsfont}
\newcommand{\ud}{\mathrm{d}}
\newcommand{\ui}{\mathrm{i}}
\newcommand{\ue}{\mathrm{e}}

\newcommand{\coloneqq}{:=}

\newcommand{\gz}{{\mathbb Z}}
\newcommand{\rz}{{\mathbb R}}
\newcommand{\nz}{{\mathbb N}}
\newcommand{\kz}{{\mathbb C}}
\newcommand{\eins}{\mathds{1}}
\newcommand{\lf}{\mathfrak{l}}
\newcommand{\wK}{\widetilde{K}}

\newcommand{\PBb}{P_{\ker B'}}
\newcommand{\PBs}{P_{\ker B'}^{\bot}}
\newcommand{\mdfrac}{\frac}

\newcommand{\ba}{\boldsymbol{a}}
\newcommand{\bb}{\boldsymbol{b}}
\newcommand{\ap}{\boldsymbol{\alpha}}
\newcommand{\bp}{\boldsymbol{\beta}}
\newcommand{\tD}{\widetilde{D}_{(\ba\bb)}}
\newcommand{\D}{D_{\boldsymbol{(\ba\bb)}}}
\newcommand{\Ps}{\boldsymbol{[\psi]}}

\DeclareMathOperator{\im}{Im}
\DeclareMathOperator{\re}{Re}
\DeclareMathOperator{\trr}{Tr}
\DeclareMathOperator{\mtr}{tr}
\DeclareMathOperator{\ran}{ran}

\DeclareMathOperator{\arth}{artanh}

\DeclareMathOperator{\Mat}{Mat}
\DeclareMathOperator{\rank}{rank}
\DeclareMathOperator{\ar}{area}

\DeclareMathOperator{\spa}{span}
\DeclareMathOperator{\mod}{mod}

\newtheorem{theorem}{Theorem}[section]

\newtheorem{prop}[theorem]{Proposition}

\newtheorem{defn}[theorem]{Definition}

\newtheorem{ex}[theorem]{Example}
\begin{document}
\title[The Berry-Keating operator on $L^2\left(\rz_>,\ud x\right)$ and on compact graphs]{The Berry-Keating operator on $L^2\left(\rz_>,\ud x\right)$ and on compact quantum graphs with general self-adjoint realizations}
\author{Sebastian Endres}
\address{Institut f{\"u}r Theoretische Physik, Universit{\"a}t Ulm\newline Albert-Einstein-Allee 11, 89081 Ulm, Germany}
\ead{sebastian.endres@uni-ulm.de}
\author{Frank Steiner}
\address{Institut f{\"u}r Theoretische Physik, Universit{\"a}t Ulm\newline Albert-Einstein-Allee 11, 89081 Ulm, Germany}
\ead{frank.steiner@uni-ulm.de}
\begin{abstract}
The Berry-Keating operator $H_{\mathrm{BK}}:= -\ui\hbar\left(x\frac{\ud\phantom{x}}{\ud x}+\frac{1}{2}\right)$ [M. V. Berry and J. P. Keating, SIAM Rev. 41 (1999) 236] governing the Schrödinger dynamics is discussed in the Hilbert space $L^2\left(\rz_>,\ud x\right)$ and on compact quantum graphs. It is proved that the spectrum of $H_{\mathrm{BK}}$ defined on $L^2\left(\rz_>,\ud x\right)$ is purely continuous and thus this quantization of $H_{\mathrm{BK}}$ cannot yield the hypothetical Hilbert-Polya operator possessing as eigenvalues the nontrivial zeros of the Riemann zeta function. A complete classification of all self-adjoint extensions of $H_{\mathrm{BK}}$ acting on compact quantum graphs is given together with the corresponding secular equation in form of a determinant whose zeros determine the discrete spectrum of $H_{\mathrm{BK}}$. In addition, an exact trace formula and the Weyl asymptotics of the eigenvalue counting function are derived. Furthermore, we introduce the ``squared'' Berry-Keating operator $H_{\mathrm{BK}}^2:= -x^2\frac{\ud^2\phantom{x}}{\ud x^2}-2x\frac{\ud\phantom{x}}{\ud x}-\frac{1}{4}$  which is a special case of the Black-Scholes operator used in financial theory of option pricing. Again, all self-adjoint extensions, the corresponding secular equation, the trace formula and the Weyl asymptotics are derived for $H_{\mathrm{BK}}^2$ on compact quantum graphs. While the spectra of both $H_{\mathrm{BK}}$ and $H_{\mathrm{BK}}^2$ on any compact quantum graph are discrete, their Weyl asymptotics demonstrate that neither $H_{\mathrm{BK}}$ nor $H_{\mathrm{BK}}^2$ can yield as eigenvalues the nontrivial Riemann zeros. Some simple examples are worked out in detail.
\phantom{\cite{Weil,Weil:1999,Weil:1952}}
\end{abstract}
\submitto{\JPA}
\maketitle
\section{Introduction: The hypothetical Hilbert-Polya operator}
\label{200}
There is an old idea, usually attributed to 
Hilbert
\cite{Weil} and Polya
\cite{Od}
that the nontrivial (i.e.\ complex) zeros $s_n$ of the Riemann zeta function $\zeta(s)$ have a spectral interpretation. Writing $s_n:= \frac{1}{2}-\ui \tau_n$, the Riemann hypothesis states that the nonimaginary solutions $\tau_n$ of $\zeta(\frac{1}{2}-\ui \tau_n)=0$ are real, that is the nontrivial zeros $s_n$ lie on the critical line $\re s=\frac{1}{2}$. The Hilbert-Polya approach towards a proof of the Riemann hypothesis consists in finding a Hilbert space $\mathcal{H}$ and a self-adjoint operator $H$ in $\mathcal{H}$ whose discrete spectrum is exactly given by the nontrivial zeros $\tau_n=\ui\left(s_n-\frac{1}{2}\right)$.

Around 1950, Selberg 
\cite{Selberg:1956}
introduced his zeta function $Z(s)$ in analogy with $\zeta(s)$ and with the intention to shed some light on the nontrivial Riemann zeros and the Riemann hypothesis. He noticed the striking similarity between his famous trace formula for the Laplace-Beltrami operator on e.g.\ compact Riemannian manifolds and the explicit formulae of number theory, whose most general form is Weil's explicit formula 
\cite{Weil:1952}. The nontrivial zeros of the Selberg zeta function $Z(s)$ fulfil the analogue of Riemann's hypothesis and appear in the spectral side of the trace formula being directly related to the  spectrum of the Laplacian. The other side of the trace formula has a purely geometrical interpretation, since it is given by a sum over the length spectrum of the closed geodesics (periodic orbits) of the geodesic flow, i.e.\ the free motion of a point particle on a given hyperbolic manifold. This system was already studied by Hadamard
\cite{Hardamrd:1898,Hardamrd:1898b}
in 1898 and has played an important role in the development of ergodic theory ever since. Hadamard proved that all trajectories in this system are unstable and that neighbouring trajectories diverge in time at an exponential rate, the most striking property of deterministic chaos.

In 1980, Gutzwiller
\cite{Gutzwiller:1980}
drew attention to this system as a prototype example of quantum chaos by identifying the Laplacian on hyperbolic manifolds with the Schrödinger operator in quantum mechanics. In this way he related the nontrivial zeros of the Selberg zeta function to the quantum energies of a dynamical system whose classical trajectories  are chaotic. Furthermore, he realized that the Selberg trace formula is an exact version of his trace formula, the celebrated Gutzwiller trace formula 
\cite{Gutzwiller:1971},
which holds for general quantum systems with a chaotic classical counterpart, but in this case only approximately, i.e.\ in the so-called semiclassical limit where Planck's constant $\hbar$ approaches zero.

In 1985, Berry
\cite{Berry:1986}
emphasized that the search for the hypothetical Hilbert-Polya operator in terms of a Schrödinger operator obtained from the quantization of a classically chaotic system might be a fruitful route to proving the Riemann hypothesis. He discussed in detail the properties of this operator that are suggested by the quantum analogy. Prompted by a paper written by Connes
\cite{Connes:1996}
(see also
\cite{Connes:1999}),
who devised a self-adjoint operator (Perron-Frobenius) of a classical dynamical system together with a classical (Lefschetz) trace formula in noncommutative geometry, Berry and Keating
\cite{Berry:1999b,Berry:1999}
speculated that the conjectured Hilbert-Polya operator might be some quantization of the extraordinarily simple classical Hamiltonian function $H_{\mathrm{cl}}(x,p)$ of a single coordinate $x$ and its conjugate momentum $p$:
\begin{equation}
\label{201}
H_{\mathrm{cl}}(x,p):= xp.
\end{equation}

Inspired by \cite{Smilansky:1997,KottosSmilansky:1998}, Berry and Keating \cite{Berry:1999b} suggested to investigate quantum graph models of \eref{201}, in particular the spectrum of these operators. One of the first researchers who dealt with differential operators on graphs was Roth \cite{Roth:1983} who derived a trace formula for the heat kernel of the Laplacian with Kirchhoff boundary conditions. Von Below \cite{Below:1985} considered the heat equation on graphs and derived a characteristic equation for the eigenvalues of the weighted Laplacian on graphs. Some physical quantum graph models were considered by Exner and \v{S}eba \cite{Exner:1989} who discussed i.a. the scattering problem for a free quantum particle on a star graph. A method to approximate mesoscopic systems like thin branching systems by quantum graphs was discussed by Exner and Post \cite{Post:2005} and Post \cite{Post:2006}. Carlson \cite{Carlson:2006}  used semigroups on graphs to simulate the blood flow in the human arterial system. Kottos ans Smilansky \cite{Smilansky:1997,KottosSmilansky:1998} introduced quantum graphs as a model for quantum chaos.    

In this paper, we study the quantization of the classical Berry-Keating Hamiltonian (\ref{201}) in the Hilbert space $L^2(\rz_>,\ud x)$ and on compact quantum graphs and give a complete classification of the self-adjoint realizations of the corresponding Berry-Keating operator. In addition, we also study the quantization of the corresponding ``squared'' operator. It turns out that no self-adjoint realization of \eref{201} exists which yields as eigenvalues the Riemann zeros. 

\section{Classical dynamics and quantization of the Berry-Keating operator}
\label{26}
Let us consider the classical dynamics of a particle moving on the real line $\rz$ generated by the Berry-Keating Hamiltonian (\ref{201}) with corresponding phase space $\mathcal{P}:(x,p)\in\rz\times\rz$. The classical time evolution (Hamiltonian flow) is governed by Hamilton's equations
\begin{equation}
\label{3}
\dot{x}(t) = \frac{\partial H_{\mathrm{cl}}}{\partial p}=x(t) \quad \mbox{and} \quad \dot{p}(t) = -\frac{\partial H_{\mathrm{cl}}}{\partial x}=-p(t).
\end{equation}
Starting at time $t=0$ at an arbitrary point $(x_0,p_0)\in\mathcal{P}$ in phase space, the unique solutions are 
\cite{Berry:1999b}
\begin{equation}
\label{4}
x(t)=x_0\ue^{t} \quad \mbox{and} \quad p(t)=p_0\ue^{-t}.
\end{equation}
Obviously, the point $(0,0)\in\mathcal{P}$ is an unstable point. We note that the Hamiltonian (\ref{201}) is time independent corresponding to the conserved ``energy'' $E:=H_{\mathrm{cl}}(x(t),p(t))=x_0p_0\in\rz$, and thus the particle moves in $\mathcal{P}$ on the ``energy surface'' (hyperbola) $xp=E$. Obviously, the classical motion is unbounded. Therefore, Berry and Keating 
\cite{Berry:1999b,Berry:1999}
introduced some regularization procedures, leading to a truncation of phase space, which we shall discuss below, but first we would like to discuss quantum mechanics.

Quantization of the classical system requires to choose a Hilbert space $\mathcal{H}$ and to replace the classical Hamiltonian (\ref{201}) by a self-adjoint operator $H$ in $\mathcal{H}$. With the standard choice $\mathcal{H}\coloneqq L^2(\rz,\ud x)$, the simplest operator corresponding to (\ref{201}) is obtained by Weyl ordering of the coordinate operator $x$ (acting by multiplication) and the momentum operator $p=-\ui\hbar\mdfrac{\ud\phantom{x}}{\ud x}$ (acting by differentiation) leading to the Berry-Keating operator
\cite{Berry:1999b,Berry:1999}
\begin{equation}
\label{29}
H_{\mathrm{BK}}\coloneqq \frac{1}{2}(xp+px)=-\ui\hbar\left(x\frac{\ud\phantom{x}}{\ud x}+\frac{1}{2}\right),
\end{equation}
and the Schrödinger equation
\begin{equation}
\label{300}
\ui \hbar \frac{\partial \Psi(x,t)}{\partial t} = H_{\mathrm{BK}}\Psi(x,t).
\end{equation}
As was to be expected from our discussion of the classical motion, the operator $H_{\mathrm{BK}}$ is unbounded and does not have a discrete spectrum corresponding to bound states, but rather has a continuous spectrum $\lambda\in\rz$ corresponding to scattering states obtained by solving the eigenvalue problem 
\begin{equation}
\label{301}
H_{\mathrm{BK}}\psi(x)=\lambda\psi(x).
\end{equation}
Writing $\lambda=\hbar k$, $k\in\rz$, Planck's constant drops out from (\ref{301}), and the eigenvalue problem reads ($s\coloneqq -\frac{1}{2}+\ui k$)
\begin{equation}
\label{302}
x\frac{\ud \phi_s(x)}{\ud x}=s\phi_s(x).
\end{equation}
For $x\in\rz$, (\ref{302}) possesses the general solution 
\begin{equation}
\label{7758}
\phi_s(x)=c_1x_+^s+c_2x_-^s,
\end{equation}
where $x_{\pm}^s$ denote the generalized functions (see e.g. \cite[p.\,87]{Gelfand:1960})
\begin{equation}
\label{303}
x_+^s\coloneqq
\cases{0&for $x\leq0$\\
x^s&for $x>0$\\}
\quad \mbox{and} \quad
x_-^s\coloneqq 
\cases{|x|^s&for $x<0$\\
0&for $x\geq0$\\},
\end{equation}
which is well defined for $\re s>-1$. In 
\cite{Berry:1999b},
Berry and Keating studied as a special case the simplest choice for the continuation of the eigenfunctions across the singularity at $x=0$ by considering the even eigenfunctions ($c_1=c_2=c$) $\phi_{s}^{\mathrm{even}}(x)=c|x|^s$.

Let us discuss in more detail the case that the quantum dynamics takes place on the positive half-line $x\in\rz_>$. Then $H_{\mathrm{BK}}$ acting on $\mathcal{D}(\rz_>)$, the set of infinitely continuous differentiable functions with compact support on $\rz_>$, is essentially self-adjoint (see e.g.\ \cite{Reed:1975} [both deficiency indices are equal to zero]). Therefore, the closure of this operator is self-adjoint. The general solution of the time-independent Schrödinger equation (\ref{301}) is then given by 
\begin{equation}
\label{360}
\psi_k(x)\coloneqq \frac{1}{\sqrt{2\pi}}x_+^{-\frac{1}{2}+\ui k} \quad \mbox{with} \quad k\in\rz,
\end{equation}
which is obviously not in $L^2(\rz_>, \ud x)$ and satisfies the orthonormality relation (in a distributional sense)
\begin{equation}
\label{361}
\left<\psi_k\mid\psi_{k'}\right>\coloneqq \int\limits_0^{\infty}\overline{\psi}_k(x)\psi_{k'}(x)\ud x=\delta(k-k')
\end{equation}
and the completeness relation 
\begin{equation}
\label{362}
\int\limits_{-\infty}^{\infty}\psi_{k}(x)\overline{\psi}_k(x')\ud k=\delta(x-x').
\end{equation}
Thus, we have for any $\phi\in L^2(\rz_>,\ud x)$ the spectral decomposition
\begin{equation}
\label{305}
\phi(x)=\int\limits_{-\infty}^{\infty}A(k)\psi_k(x)\ud k
\end{equation}
with 
\begin{equation}
\label{480}
A(k)\coloneqq \left<\psi_k\mid\phi\right> =\int\limits_{0}^{\infty}\overline{\psi}_k(x)\phi(x)\ud x
\end{equation}
and (assuming $\left<\phi\mid\phi\right>=1$)
\begin{equation}
\label{306}
\int\limits_{-\infty}^{\infty}|A(k)|^2\ud k = 1.
\end{equation}
Forming a general wave packet with a given amplitude $A(k)$ satisfying (\ref{305}) and (\ref{306}), one obtains ($x\in\rz_>$)
\begin{equation}
\label{307}
\phi(x)=\sqrt{\frac{2\pi}{x}}\hat{A}(\ln x),
\end{equation}
where $\hat{A}$ denotes the Fourier transform of $A$
\begin{equation}
\label{308}
\hat{A}(y)\coloneqq \frac{1}{2\pi}\int\limits_{-\infty}^{\infty}A(k)\ue^{\ui ky}\ud k.
\end{equation}
Let us also mention an alternative to the spectral decomposition (\ref{305}) for the Berry-Keating operator. Defining the Mellin transform $\check{\phi}$ of $\phi$ by
\begin{equation}
\label{2000}
\check{\phi}(s)\coloneqq \int\limits_{0}^{\infty}x^{s-1}\phi(x) \ud x,
\end{equation}
we obtain for the wave number amplitude $A(k)$ (see (\ref{480}))
\begin{equation}
\label{2001}
A(k)= \frac{1}{\sqrt{2\pi}}\check{\phi}\Bigl(\frac{1}{2}-\ui k\Bigr),
\end{equation}
from which $\phi(x)$ can be recovered by the inverse Mellin transform (convergent at least in mean square, see e.g. \cite[p.\,94]{Titchmarsh:1948})
\begin{equation}
\label{2002}
\eqalign{
\phi(x) &= \frac{1}{2\pi\ui}\int\limits_{\frac{1}{2}-\ui\infty}^{\frac{1}{2}+\ui \infty}\check{\phi}(s)x^{-s}\ud s = \frac{1}{2\pi}\int\limits_{-\infty}^{\infty}\check{\phi}\Bigl(\frac{1}{2}-\ui k\Bigr)x^{-\frac{1}{2}+\ui k} \ud k \\
&= \int\limits_{-\infty}^{\infty}A(k)\psi_k(x)\ud k 
}
\end{equation}
in agreement with (\ref{305}).

The unitary group 
\begin{equation}
\label{309}
U(t)\coloneqq \exp\Bigl(-\frac{\ui t}{\hbar}H_{\mathrm{BK}}\Bigr)=\ue^{-\frac{t}{2}}\ue^{-tD}
\end{equation}
generated by the Berry-Keating operator (\ref{29}) acts on functions $\phi\in\mathcal{H}$ as (see i.a \cite[p.\,365]{Weidmann:2003b})
\begin{equation}
\label{310}
(U(t)\phi)(x)=\ue^{-\frac{t}{2}}\phi\left(\ue^{-t}x\right).
\end{equation}
Here we have used the relation $H_{\mathrm{BK}}=-\ui\hbar\left(D+\frac{1}{2}\right)$, where $D\coloneqq x\mdfrac{\ud\phantom{x}}{\ud x}$ is the generator of scaling transformations (dilations). Let us mention that the operator $D$ has been discussed by Arendt \cite{Arendt:1994,Arendt:1995,Arendt:2002}, where $A_p\coloneqq -D$ is considered as the generator of a semigroup on e.g.\ $L^p(\rz_>,\ud x)$ ($1\leq p<\infty$) with Dirichlet and Neumann boundary conditions.

On the other hand, the action of the unitary operator $U(t)$ on eigenfunctions $\psi$ of $H_{\mathrm{BK}}$ gives according to (\ref{301})
\begin{equation}
\label{8680}
(U(t)\psi)(x)=\ue^{-\ui\frac{\lambda}{\hbar}t}\psi(x),
\end{equation}
which in turn leads with (\ref{310}), $\lambda=\hbar k$, $s=-\frac{1}{2}+\ui k$ and $\kappa\coloneqq \ue^{-t}>0$ ($t<\infty$) to
\begin{equation}
\label{311}
\psi(\kappa x)=\kappa^s \psi(x).
\end{equation}
This shows that an eigenfunction $\psi$ of $H_{\mathrm{BK}}$ must be a homogeneous function  with (complex) degree $s=-\frac{1}{2}+\ui k$. Differentiation of (\ref{311}) with respect to $\kappa$ and then setting $\kappa=1$ leads back to the eigenvalue problem (\ref{302}) which possesses for $x\in \rz_>$ the unique solution (\ref{360})

For the (retarded) integral kernel $K_{\mathrm{BK}}(x,x_0;t)$ of the time-evolution operator $U(t)$ one obtains ($x,x_0\in\rz_>$; $\Theta(t)$ is the Heaviside step function)
\begin{equation}
\label{312}
K_{\mathrm{BK}}(x,x_0;t)= \ue^{-\frac{t}{2}}\delta\left(x_0-x\ue^{-t}\right)\Theta(t).
\end{equation}
We observe that the quantum mechanical time evolution follows in the configuration space exactly the classical trajectory (\ref{4}). Starting at time $t=0$ with the initial wave function $\phi\in L^2(\rz_>,\ud x)$, one obtains with (\ref{312}) the wave function $\psi(x,t)$ at a later time $t>0$
\begin{equation}
\label{313}
\eqalign{
\psi(x,t) &=\int\limits_{0}^{\infty}K_{\mathrm{BK}}(x,x_0;t)\phi(x_0)\ud x_0 \\
          &=\ue^{-\frac{t}{2}}\phi\left(\ue^{-t}x\right) 
}
\end{equation}
in complete agreement with (\ref{310}). We also give the result for the resolvent kernel (outgoing Green's function [a small positive imaginary part ($\epsilon>0$) has been added to $\lambda=\hbar k$]), see e.g.\ \cite[p.\,26]{Steiner:1998},
\begin{equation}
\label{314}
\eqalign{
G_{\mathrm{BK}}(x,x_0;\lambda) &\coloneqq \frac{\ui}{\hbar}\int\limits_{0}^{\infty}\ue^{\frac{\ui}{\hbar}(\lambda+\ui\epsilon)t} K_{\mathrm{BK}}(x,x_0;t)\ud t \\
          &=\int\limits_{-\infty}^{\infty}\frac{\psi_{k'}(x)\overline{\psi}_{k'}(x_0)}{\hbar k'-\lambda -\ui \epsilon}\ud k' \\
          &=\frac{2\pi\ui}{\hbar}\psi_k(x)\overline{\psi}_k(x_0)\Theta(x-x_0),
}
\end{equation}
which satisfies the inhomogeneous time-independent Schrödinger equation (see (\ref{301}))
\begin{equation}
\label{315}
\left(H_{BK,x}-\hbar k\right)G_{\mathrm{BK}}(x,x_0;\hbar k)=\delta(x-x_0).
\end{equation}

Since the operator (\ref{29}) acting in the Hilbert space $L^2(\rz,\ud x)$ respectively $L^2(\rz_>,\ud x)$ has only a continuous spectrum, it cannot be considered (with the above realization) as a candidate for the hypothetical Hilbert-Polya operator. Thus, there remains the task to find another Hilbert space for which the quantization of the classical Hamiltonian (\ref{201}) possesses a discrete spectrum. Perhaps the required space is a quantum graph, with $xp$ acting on edges between vertices, a possibility already mentioned by Berry and Keating
\cite{Berry:1999b}. It is the purpose of our paper to discuss the self-adjoint realizations on compact quantum graphs and in a forthcoming paper 
\cite{SteinerEndres:2008} we study a noncompact quantum graph.
\section{Semiclassical regularization of the Berry-Keating operator}
\label{599}
Before we come to an investigation of quantum graphs, we would like to discuss an alternative and very interesting approach also put forward by Berry and Keating
\cite{Berry:1999b}
(see also Connes 
\cite{Connes:1996,Connes:1999})
which is based on semiclassical arguments. It is well known that the number of quantum levels with energy less than $E$, the counting function $N(E)$, is for any classical bounded Hamiltonian $H_{\mathrm{cl}}(x,p)$ in one dimension given by (see e.g.\ \cite{Gutzwiller:1990}))
\begin{equation}
\label{500}
N(E)= \frac{1}{2\pi\hbar}\ar(E)\bigl(1+\Or(\hbar)\bigr),
\end{equation}
where
\begin{equation}
\label{501}
\ar(E)\coloneqq \int\limits_{\mathcal{P}}\ud x \ud p \ \Theta\bigl(E-H_{\mathrm{cl}}(x,p)\bigr)
\end{equation}
is the phase-space area under the contour $H_{\mathrm{cl}}(x,p)=E$. Obviously, there is a problem if this formula is applied to the Hamiltonian (\ref{201}), since the classical motion is not bounded, so that $\ar(E)$ is infinite. Therefore, Berry and Keating 
\cite{Berry:1999b}
proposed to regularize the system by a suitable truncation of phase space in such a way that $\ar(E)$ becomes finite.

The regularization proposed by Berry and Keating 
\cite{Berry:1999b}
is to truncate $x$ and $p$ by considering the ``regularized phase space'' $\mathcal{P}_{\mathrm{reg}}\coloneqq (l_x,\infty)\times(l_p,\infty)$ together with the semiclassical condition $l_xl_p=2\pi\hbar$. This truncation cuts off not only the ``small'' coordinate $x\leq l_x$ respectively momentum values $p\leq l_p$, but it leads for a given ``energy'' $E>0$ also to a cut off at the ``large'' values $x=\frac{E}{l_p}$ respectively $p=\frac{E}{l_x}$ since $E=H_{\mathrm{cl}}(x,p)$ holds. Without specifying the behaviour of the classical motion at the end points of the trajectories, we follow Berry and Keating and obtain from (\ref{500}) and (\ref{501})
\begin{equation}
\label{502}
\eqalign{
N(E) & = \frac{1}{2\pi\hbar}\left(\int\limits_{l_x}^{\frac{E}{l_p}}\frac{E}{x}\ud x-l_p\left(\frac{E}{l_p}-l_x\right)\right)\bigl(1+\Or(\hbar)\bigr)\\
     & =\frac{1}{2\pi\hbar}E\left(\ln\left(\frac{E}{2\pi\hbar}\right)-1\right)+1+\ldots.
}
\end{equation}
Setting $\hbar=1$ together with a modification of $N(E)$ by adding $-\frac{1}{8}$ to the right-hand side of (\ref{502}) which was suggested by Berry and Keating \cite{Berry:1999b,Berry:1999} in order to take into account the Maslov index, we arrive at the leading asymptotics of the counting function of the nontrivial zeros of the Riemann zeta function (Riemann- von Mangoldt formula)
\begin{equation}
\label{2}
N(E)=\frac{E}{2\pi}\ln\left(\frac{E}{2\pi}\right)-\frac{E}{2\pi}+\frac{7}{8}+\Or\left(\ln E\right).
\end{equation}
Following the argumentation of Berry and Keating \cite{Berry:1999b,Berry:1999} for the modification of $N(E)$, we get for the corresponding Maslov index $\mu=-\frac{1}{2}$. This seems at first somewhat strange since there is no magnetic flux or spinning particle given and, therefore, the Maslov index should be an integer number as in the case of ``normal'' quantum systems like the harmonic oscillator. We want to mention that there is actually no rigorous argument for the choice of the Maslov index (correction) simply by the fact that so far we have not yet imposed any boundary conditions on the operator, and in the corresponding classical description there is therefore a lack of jump or scattering condition at the end points of the trajectories. The scattering conditions in section \ref{120} (example \ref{50000}) could provide a possible remedy for the above mentioned discrepancy of the Maslov index with respect to ``normal'' systems. Furthermore, there is only one possibility in the classical case for the behaviour of the particle at the end point of the trajectory if one wants to preserve the constancy of the Hamiltonian for all time: the particle must jump from the point $\bigl(\frac{E}{l_p},l_p\bigr)$ to the point $\bigl(l_x,\frac{E}{l_x}\bigr)$ in phase space, which corresponds to a kind of ring-system (one-dimensional torus with the topology of $S^1$) in the configuration space.
\section{Classical dynamics and quantization of the ``squared'' Berry-Keating operator}
\label{7680}
In order to allow some kind of reflection at the end points of the trajectories, we shall also consider the classical Hamiltonian
\begin{equation}
\label{31}
\widetilde{H}_{\mathrm{cl}}(x,p)\coloneqq x^2p^2,
\end{equation}
which is the square of the Berry-Keating Hamiltonian (\ref{201}). Note that (\ref{31}) can be derived from the Lagrangian 
\begin{equation}
\label{2010}
L(x,\dot{x})=\frac{1}{4}\left(\frac{\dot{x}}{x}\right)^2
\end{equation}
and that Hamilton's equations do not decouple in this case as in (\ref{3}). In fact, one obtains
\begin{equation}
\label{2011}
\dot{x}(t) = \frac{\partial \widetilde{H}_{\mathrm{cl}}}{\partial p}=2x^2p(t) \quad \mbox{and} \quad \dot{p}(t) = -\frac{\partial \widetilde{H}_{\mathrm{cl}}}{\partial x}=-2xp^2(t),
\end{equation}
and the solutions are 
\begin{equation}
\label{5}
x(t)=x_0\ue^{2x_0p_0t} \quad \mbox{and} \quad p(t)=p_0\ue^{-2x_0p_0t}.
\end{equation}
If one broadens the phase space to
\begin{equation}
\label{2015}
\mathcal{P}_{\mathrm{reg},b}\coloneqq (l_x,\infty)\times \bigl((l_p,\infty)\cup(-l_p,-\infty)\bigr)
\end{equation}
one now has the possibility to scatter from the end point $\bigl(\frac{E}{l_p},l_p\bigr)$ of a trajectory of the form (\ref{5}) to the end point $\bigl(\frac{E}{l_p},-l_p\bigr)$. This corresponds to a reflection on a wall like in a one-dimensional billiard system. This is one reason why we rather consider $H_{\mathrm{cl}}$ and accordingly $H_{\mathrm{BK}}$ as a momentum (operator) and $\widetilde{H}_{\mathrm{cl}}$ and respectively ${H}_{\mathrm{BK}}^2$ as an energy (operator). Further hints to this choice will follow in the sequel.

Before investigating the ``squared'' Berry-Keating operator on quantum graphs, we would like to consider this operator in the framework of standard quantum mechanics restricting ourselves, however, to the positive half-line $\rz_>$ as in the discussion of the original Berry-Keating operator in section \ref{26}. A formal calculation of $\widetilde{H}\coloneqq H_{\mathrm{BK}}^2$ gives (setting from now on $\hbar=1$):
\begin{equation}
\label{30}
H_{\mathrm{BK}}^2\coloneqq \left(-\ui\Bigl(x\frac{\ud\phantom{x}}{\ud x}+\frac{1}{2}\Bigr)\right)^2=-x^2\frac{\ud^2\phantom{x}}{\ud x^2}-2x\frac{\ud\phantom{x}}{\ud x}-\frac{1}{4}.
\end{equation}
Again as in section \ref{26}, $H_{\mathrm{BK}}^2$ acting on $\mathcal{D}(\rz_>)$ is essentially self-adjoint, and in the following we always consider the self-adjoint closure of this operator. It is worthwhile to mention that the squared operator (\ref{30}) is a special case of the famous Black-Scholes operator \cite{Black:1973,Merton:1973} introduced to determine the pricing of options in financial theory whose interesting mathematical properties have been discussed e.g.\ in \cite{Arendt:1994,Arendt:1995,Arendt:2002}.

It is easy to see that the functions $\psi_k(x)$ ($k\in\rz\setminus\{0\}$) defined in (\ref{360}) are the only eigenfunctions of $H_{\mathrm{BK}}^2$ on $\rz_>$ corresponding to the continuous spectrum $\lambda=k^2>0$. Here the eigenvalue $\lambda=0$ (respectively $k=0$) corresponds to the two generalized eigenfunctions
\begin{equation}
\label{4700}
\psi_{0,1}(x)=\frac{1}{\sqrt{2\pi}}x_+^{-\frac{1}{2}} \quad \mbox{and} \quad \psi_{0,2}(x)=\frac{1}{\sqrt{2\pi}}x_+^{-\frac{1}{2}}\ln x.
\end{equation}
An eigenvalue $\lambda=k^2>0$ possesses the two linearly independent generalized eigenfunctions $\psi_k(x)$ and $\psi_{-k}(x)$.

Introducing the (retarded) integral kernel of the time-evolution operator (unitary group) 
\begin{equation}
\label{7701}
\widetilde{U}(t)\coloneqq \ue^{-\ui tH_{\mathrm{BK}}^2}
\end{equation}
by
\begin{equation}
\label{7702}
\psi(x,t)\coloneqq \left(\widetilde{U}(t)\phi\right)(x)=:\int\limits_{\rz_>}\widetilde{K}(x,x_0;t)\phi(x_0)\ud x_0,
\end{equation}
where $\phi(x)\in L^2\left(\rz_>,\ud x\right)$ is the initial wave function at $t=0$, we obtain (cf.\ \cite[p.\,27]{Steiner:1998}) 
\begin{equation}
\label{7703}
\eqalign{
\widetilde{K}(x,x_0;t) &= \int\limits_{-\infty}^{\infty}\psi_k(x)\overline{\psi}_k(x_0)\ue^{-\ui k^2 t}\Theta(t)\ud k \\
                       &= \left(4\pi \ui txx_0\right)^{-\frac{1}{2}}\ue^{\ui\frac{\left(\ln x-\ln x_0\right)^2}{4t}}\Theta(t).
}
\end{equation}
The kernel $\widetilde{K}$ satisfies the inhomogeneous time-dependent Schrödinger equation 
\begin{equation}
\label{7073}
\eqalign{
\left(\ui \frac{\partial\phantom{t}}{\partial t}-H_{BK,x}^2\right)\widetilde{K}(x,x_0;t)=\ui \delta(x-x_0)\delta(t),
}
\end{equation}
i.e.\ it is the retarded Green's function. With (\ref{7702}), the action of $\widetilde{U}(t)$ on $\phi\in L^2\left(\rz_>,\ud x\right)$ is given by ($t>0$)
\begin{equation}
\label{7074}
\left(\widetilde{U}(t)\phi\right)(x)=\left(4\pi\ui t\right)^{-\frac{1}{2}}\int\limits_{-\infty}^{\infty}\ue^{\ui\frac{\tau^2}{4t}}\ue^{-\frac{\tau}{2}}\phi\left(\ue^{-\tau}x\right)\ud \tau,
\end{equation}
which expresses the fact that $\widetilde{U}(t)$ is a combination of the scaling transformation generated by the operator $D=x\mdfrac{\ud\phantom{x}}{\ud x}$ (see eq. (\ref{310})) and the transformation generated by the operator $T\coloneqq x^2\mdfrac{\ud^2\phantom{x}}{\ud x^2}$, since $\widetilde{U}(t)=\ue^{\ui \frac{t}{4}}\ue^{\ui tT}\ue^{2\ui t D}$. Notice that the transformation generated by $T$ reads
\begin{equation}
\label{7075}
\left(\ue^{\ui tT}\phi\right)(x)=\frac{\ue^{-\ui\frac{t}{4}}}{\left(4\pi \ui t\right)^{\frac{1}{2}}}\int\limits_{-\infty}^{\infty}\ue^{\ui\frac{\tau^2}{4t}+\frac{\tau}{2}}\phi\left(\ue^{-\tau}x\right)\ud \tau,
\end{equation}
and that the operators $D$ and $T$ commute. The resolvent kernel (outgoing Green's function) of $H_{\mathrm{BK}}^2$ is given by (see \cite[p.\,26]{Steiner:1998})
\begin{equation}
\label{7076}
\eqalign{
\widetilde{G}(x,x_0;\lambda) &\coloneqq  \ui\int\limits_{0}^{\infty}\ue^{\ui(\lambda+\ui\epsilon)t}\widetilde{K}\left(x,x_0;t\right)\ud t \\
                             &= \left(4xx_0\left(-\lambda-\ui\epsilon\right)\right)^{-\frac{1}{2}}\ue^{-{\left(-\lambda-\ui\epsilon\right)}^{\frac{1}{2}}\left|\ln x-\ln x_0\right|},
}
\end{equation}
which shows that $\widetilde{G}$ has a cut on the positive real axis in the complex $\lambda$-plane (if $\sqrt{z}$ is defined with a cut on the negative real axis in the $z$-plane). With $k\coloneqq \sqrt{\lambda}>0$ one obtains ($x,x_0\in\rz_>$)
\begin{equation}
\label{7077}
\eqalign{
\widetilde{G}\left(x,x_0;k^2\right) &= \frac{\ui}{2k\sqrt{xx_0}}\ue^{\ui k\left|\ln x-\ln x_0\right|} \\
                                    &= \frac{\ui \pi}{k}
                                    \cases{
                                    \psi_k(x)\overline{\psi}_k(x_0) & for $x\geq x_0$\\
                                    \overline{\psi}_k(x)\psi_k(x_0) & for $x<x_0$\\
                                    }
}
\end{equation}
in agreement with the general form of the Green's function of a Sturm-Liouville operator (see e.g.\ \cite[p.\,112]{Terras:1985}).
\section{Semiclassical estimate for the eigenvalue counting function of the ``squared'' Berry-Keating operator}
\label{10599}
Using again the semiclassical formula (\ref{500}) and the truncation of phase space as discussed in section \ref{599}, we obtain for the counting function in the quadratic case
\begin{equation}
\label{505}
\eqalign{
N(E)&=\frac{1}{2\pi\hbar}\,2\left(\int\limits_{l_x}^{\frac{\sqrt{E}}{l_p}}\frac{\sqrt{E}}{x}\ud x-l_p\left(\frac{\sqrt{E}}{l_p}-l_x\right)\right)\bigl(1+\Or(\hbar)\bigr)\\
    &=2\left[\frac{k}{2\pi}\ln\left(\frac{k}{2\pi}\right)-\frac{k}{2\pi}+\frac{7}{8}\right]+\ldots,
}
\end{equation}
where we have included the same Maslov index correction as in section \ref{599}. Furthermore, we have introduced the ``wave number'' $k$, $E=:\hbar^2k^2$, and have used $l_xl_p=2\pi\hbar$. We note that in this case we obtain twice the counting function of the Riemann zeros (for which only those with positive imaginary part are counted), since each energy value $E$ comes with two values $\pm k$. Notice, that in this case the Riemann zeros are not interpreted as ``energies'' but rather as ``momenta'' $\hbar k$ respectively ``wave numbers'' $k$. Formula (\ref{505}) agrees with the well-known universal law that $N(E)$ for a bounded system in $d$ dimensions grows asymptotically as $N(E)=\Or\bigl(E^{\frac{d}{2}}\bigr)$, and thus for a one-dimensional system one expects $N(E)=\Or\bigl(\sqrt{E}\bigr)=\Or(k)$, eventually modified by a factor $\ln\bigl(\sqrt{E}\bigr)$.
\section{Compact graphs}
\label{7}

We shall present a short overview on compact graphs using the notations of \cite{KostrykinSchrader:1999} and \cite{BE:2008}.

A compact graph $\Gamma=(\mathcal{V},\mathcal{E},\mathbf{I})$ is a finite set of vertices $\mathcal{V}=(v_1,\ldots,v_V)$ and a finite set of edges $\mathcal{E}=(e_1,\ldots,e_E)$. Here we have defined $E\coloneqq |\mathcal{E}|$ and $V\coloneqq |\mathcal{V}|$ for the total number of edges and vertices, respectively. Each vertex $v\in \mathcal{V}$ is at least connected with one element $\tilde{v}\in \mathcal{V}$ by some edge $e\in \mathcal{E}$, where $v=\tilde{v}$ is allowed. Furthermore, each edge $e\in \mathcal{E}$ connects two vertices $v$ and $\tilde{v}$ in $\mathcal{V}$, again $v=\tilde{v}$ is possible. The topology of the graph is given by these relations of the edges and the vertices. Each edge $e$ is assigned an interval $I_e=[a_e,b_e]$ with $0<a_e<b_e<\infty$. The set of all intervals is denoted by $\mathbf{I}$. We remark that the choice of the starting point $a_e$ and the final point $b_e$ of the edge $e$ is arbitrary and there is no orientation of the graph assumed. We denote two edges as adjacent iff they share at least one vertex as endpoint. We need the notion of a path and of a periodic orbit of the graph. We slightly differ from the definition in \cite{BE:2008} for further convenience. A path $p(w,z)\coloneqq \left((e_i)_{i=1}^{n},w,z\right)$ is a set of a finite sequence of edges $(e_i)_{i=1}^n$ where the points $w,z\in\mathbf{I}$ denote the starting and final points of the path. Furthermore, it is required that
\begin{itemize}
\item the edges $e_i$ and $e_{i+1}$ are adjacent,
\item the point $w$ must be an element of $I_{1}$ and $z$ must be an element of $I_{n}$.
\end{itemize}
The case $w=z$ is admissible and corresponds to a closed path. In \cite{BE:2008} or \cite{KottosSmilansky:1998} only the first item is required for a closed path at which we set $e_1=e_{n+1}$. We shall call this case a closed orbit. Especially, a closed orbit is only characterized by a sequence of edges $(e_i)_{i=1}^n$. For the definition of a periodic orbit $\gamma$, we shall keep with the usual definition as in \cite{BE:2008}, then $\gamma$ is an equivalence class of closed orbits and can be characterized by a representative $\gamma=(e_i)_{i=1}^n$. The set of all periodic orbits is denoted by $\mathfrak{P}$.
We could then equip the graph with a metric structure in an obvious way like in \cite{BE:2008}. Especially, this would mean that the length of the edge $e_i$ will be $l_i=b_i-a_i$. However, here we take another choice for the lengths and the metric structure of the graph. We define the length $\mathfrak{l}_{p}(w,z)$ of the path $p(w,z)\coloneqq \left((e_i)_{i=1}^{n},w,z\right)$ as follows. If $n\geq2$ we denote by $y_1$ and $y_n$ the endpoints of the intervals $I_1$ and $I_n$ corresponding to the shared vertices of the edges $e_1,e_2$ and $e_{n-1},e_n$. In particular this means that $y_1$ is identical with $a_1$ or $b_1$ and $y_n$ is identical with $a_n$ or $b_n$. Then the length $\mathfrak{l}_{p}(w,z)$ is defined as
\begin{equation}
\label{80}
\mathfrak{l}_{p}(w,z)\coloneqq \left|\ln\left(\frac{y_1}{w}\right)\right|+\sum\limits_{i=2}^{n-1}\ln\left(\frac{b_i}{a_i}\right)+\left|\ln\left(\frac{y_n}{z}\right)\right|.
\end{equation}
Similarly, if $n=2$ ($y_1=y_n:=y$) respectively $n=1$ we define
\begin{equation}
\label{80a}
\mathfrak{l}_{p}(w,z)\coloneqq \left|\ln\left(\frac{y}{w}\right)\right|+\left|\ln\left(\frac{y}{z}\right)\right| \quad \mbox{respectively} \quad \mathfrak{l}_{p}(w,z)\coloneqq \left|\ln\left(\frac{w}{z}\right)\right|.
\end{equation}
Furthermore, we define in a natural way the length $\mathfrak{l}_{\gamma}$ of a periodic orbit $\gamma$
\begin{equation}
\label{81}
\mathfrak{l}_{\gamma}\coloneqq \ln\left(\prod\limits_{i=1}^{n}\frac{b_i}{a_i}\right).
\end{equation}
In order to define a metric structure of the graph, we need the notion of connectedness. We define $w$ and $z$ as connected iff there exists a path $p(w,z)\coloneqq \left((e_i)_{i=1}^{n},w,z\right)$. The graph $\Gamma$ is connected iff all points of the intervals $\mathbf{I}$ are connected. Not necessarily but for convenience, we assume in the following that the graph $\Gamma$ is connected. The distance $d_{w,z}$ of two points $w$ and $z$ on the edges of the graph is defined by
\begin{equation}
\label{82}
d_{w,z}\coloneqq \min \left\{\mathfrak{l}_{p}; \ p \ \mbox{connects} \ w \ \mbox{and} \ z\right\}.
\end{equation}
We remark that this choice of the metric for the graph will correspond to a ``hyperbolic'' metric in one dimension. 
In this case the determinant of the metric tensor $g$ at the point $x$ is $(\det g)(x)=\mdfrac{1}{x^2}$. The reason for our choice of the metric will be explained in sections \ref{9} and \ref{90}.

We also need for the interpretation of the trace formula for the Berry-Keating operator $H_{\mathrm{BK}}$ in theorem \ref{11104} the notion of a directed graph in order to interpret the right side of 
(\ref{19905})
as a sum of periodic orbits. Therefore, we replace the edges by directed edges. This doesn't affect the lengths of the edges but has of course an influence on topological properties of the graph such as connectedness and on the set $\mathfrak{P}$ of periodic orbits. Since in this case there are only such paths $p(w,z)\coloneqq \left((e_i)_{i=1}^{n},w,z\right)$ allowed for which for all consecutive edges $e_i,e_{i+1}$, there exist vertices $v_{ij}$ such that the direction of $e_i$ is towards $v_{ij}$ and $e_{i+1}$ has the direction away from $v_{ij}$. Then, the definition for the periodic orbits and for connectedness for directed graphs are the same as for undirected graphs. Again, not necessarily but for convenience we always assume that the directed graph (in the case for the Berry-Keating operator $H_{\mathrm{BK}}$) is connected.
\section{The Berry-Keating operator on compact quantum graphs}
\label{9}
We define, in accordance with \cite{BE:2008}:
\begin{equation}
\label{10}
\eqalign{
\phantom{\mathcal{H}\coloneqq} C^{\infty}_0(\Gamma)& \coloneqq \bigoplus\limits_{i=1}^{E}C^{\infty}_0[a_i,b_i] \quad \mbox{and} \\ \mathcal{H}\coloneqq L^2(\Gamma)& \coloneqq \bigoplus_{i=1}^{E}L^2\left([a_i,b_i],\ud x\right)  \quad \mbox{with} \quad 0<a_i<b_i<\infty.
}
\end{equation}
This means that a function of the Hilbert space $\mathcal{H}$ is represented by an orthogonal sum of functions which are defined on the corresponding edges:
\begin{equation}
\label{32}
\psi\in L^2(\Gamma) \quad \mbox{iff} \quad \psi=\bigoplus\limits_{i=1}^E \psi_i \quad \mbox{with} \quad \psi_i\in L^2(I_i,\ud x).
\end{equation}

The first function space in (\ref{10}) will be a possible operator core for the closed Berry-Keating operator with (\ref{12}) as domain of definition. Therefore, the self-adjoint extensions of $H_{\mathrm{BK}}$ are also with respect to $C_0^{\infty}(\Gamma)$. The space $\mathcal{H}$ is a Hilbert space if we equip it with the scalar product
\begin{equation}
\label{33}
\left<\psi,\phi\right>\coloneqq \sum\limits_{j=1}^E\int\limits_{a_j}^{b_j}\overline{\psi_j(x_j)}\phi_j(x_j)\ud x_j.
\end{equation}
We then define the Berry-Keating operator on compact graphs (in the following we set $\hbar=1$):
\begin{equation}
\label{11}
H_{\mathrm{BK}}\psi\coloneqq \left(-\ui\Bigl(x\frac{\ud\phantom{x}}{\ud x}+\frac{1}{2}\Bigr)\psi_1,\ldots,-\ui\Bigl(x\frac{\ud\phantom{x}}{\ud x}+\frac{1}{2}\Bigr)\psi_{E}\right),
\end{equation}
for $\psi\in C^{\infty}_0(\Gamma)$. Since we have a compact graph $\Gamma$, multiplication by $x$ is a bounded closable operation. Thus, by perturbation arguments, see e.g.\ \cite[p.\,183]{Weidmann:2000}, we conclude that $H_{\mathrm{BK}}$ is closable since the standard momentum operator $p=-\ui\mdfrac{\ud\phantom{x}}{\ud x}$ is closable. Furthermore, we note that multiplication by the argument is also a (bounded) bijection from
\begin{equation}
\label{12}
D_{0}^1(\Gamma)\coloneqq \bigoplus\limits_{i=1}^{E}H_0^1[a_i,b_i]
\end{equation}
to itself. $H_0^1[a_i,b_i]$ is the set of absolutely continuous functions on $[a_i,b_i]$ which vanish at the endpoints of the intervals and with square integrable weak derivatives. Again, by perturbation arguments for the momentum operator, we therefore conclude that the domain of definition of the closure of $H_{\mathrm{BK}}$ is equal to (\ref{12}). Furthermore, by similar arguments the adjoint operator of $(H_{\mathrm{BK}},D_{0}^1(\Gamma))$ is given by $(H_{\mathrm{BK}},H^1(\Gamma))$ in which 
\begin{equation}
\label{41}
H^1(\Gamma)\coloneqq \bigoplus\limits_{i=1}^E H^1[a_i,b_i]
\end{equation}
is the set of absolutely continuous functions on the intervals of the graph $\Gamma$, cf.\ \cite[p.\,100]{Weidmann:2000} possessing square integrable weak derivatives.

We mention at this point that the projections of the spaces $D_0^1(\Gamma)$ and $H^1(\Gamma)$ on the intervals of the graph $\Gamma$ coincide with the corresponding Sobolev spaces, see e.g.\ \cite{Reed:1980}. The operator $(H_{\mathrm{BK}},D_{0}^1(\Gamma))$ is symmetric and it is possible to show that the deficiency indices are $(E,E)$, compare e.g.\ \cite[p.\,142]{Reed:1975}. By a proper Sobolev embedding theorem and the compactness of the graph $\Gamma$ it follows that the differential operator on $D_{0}^1(\Gamma)$ possesses a compact resolvent, see also \cite{Kuchment:2004}. Thus, by the compact resolvent theorem, cf.\ \cite[p.\,245]{Reed:1978}, and the relatively compact perturbation theorem (cf.\ \cite[p.\,113]{Reed:1978}) the operator (\ref{11}) possesses a purely discrete spectrum. 

\section{Classification of the self-adjoint extensions of the Berry-Keating operator}
\label{14} 
In order to characterize the self-adjoint extensions, we follow the ideas of \cite[p.\,138]{Reed:1975} and \cite{KostrykinSchrader:1999}, see also \cite{Everitt:2005} for a comprehensive discussion. Therefore, we define the complex symplectic form on $H^1(\Gamma)\times H^1(\Gamma)$ (cf.\ \cite[p.\,138]{Reed:1975}):
\begin{equation}
\label{13}
\left[\phi,\psi\right]_1\coloneqq \left<\phi,H_{\mathrm{BK}}^+\psi\right>_{L^2(\Gamma)}-\left<H_{\mathrm{BK}}^+\phi,\psi\right>_{L^2(\Gamma)} \quad \mbox{for} \quad \phi,\psi\in H^1(\Gamma).
\end{equation}
We call a subspace $\mathcal{X}$ $\left[\cdot,\cdot\right]_1$-symmetric, iff $\left[\phi,\psi\right]_1=0$ for all $\phi,\psi\in\mathcal{X}$. Due to the von Neumann extension theory (see e.g.\ \cite{Reed:1975}) the self-adjoint extensions are exactly the maximal $\left[\cdot,\cdot\right]_1$-symmetric subspaces of $H^1(\Gamma)$. We follow the approach by Kostrykin and Schrader \cite{KostrykinSchrader:1999} to classify these extensions.
By a proper Sobolev embedding theorem we can define the boundary value $\mathrm{bv}$ (an element of $\kz^{2E}$) of $\psi\in H^1(\Gamma)$:
\begin{equation}
\label{15}
\Psi_{\mathrm{bv}}\coloneqq \bigl(\psi_1(a_1),\ldots,\psi_E(a_E),\psi_{1}(b_1),\ldots,\psi_{E}(b_{E})\bigr)^T \quad \mbox{for} \quad \psi\in H^1(\Gamma) .
\end{equation}
For convenience, we also define:
\begin{equation}
\label{16}
I_{\pm} \coloneqq 
\left(
\begin{array}{cc}
\eins_{E\times E} & 0 \\
0 & -\eins_{E\times E}\\
\end{array}
\right),
\quad 
D_{(\ba\bb)} \coloneqq 
\left(
\begin{array}{cc}
\ba & 0 \\
0 & \bb
\end{array}
\right),
\end{equation}
with (no summation over $i$)
\begin{equation}
\label{60}
\ba_{ij}\coloneqq \delta_{ij}a_i \quad \mbox{and} \quad \bb_{ij}\coloneqq \delta_{ij}b_i \quad \mbox{for}  \quad 0\leq i,j\leq E 
\end{equation}
and
\begin{equation}
\label{21}
J\coloneqq 
\left(
\begin{array}{cc}
0 & \eins_{E\times E} \\
-\eins_{E\times E} & 0
\end{array}
\right),
\qquad  U\coloneqq \frac{1}{\sqrt{2}}
\left(
\begin{array}{cc}
\ui \eins_{E\times E} & \eins_{E\times E} \\
-\eins_{E\times E} & -\ui\eins_{E\times E}
\end{array}
\right).
\end{equation}
By a simple calculation  we obtain the identity:
\begin{equation}
\label{3001}
U\left(\ui I_{\pm}\right)U^+=J.
\end{equation}
Thus, we obtain for (\ref{13}) by an integration by parts using the unitarity of $U$:
\begin{equation}
\label{19}
\eqalign{
\left[\psi,\phi\right]_1 &= \left<\Phi_{\mathrm{bv}},\ui I_{\pm}D_{(\ba\bb)}\Psi_{\mathrm{bv}}\right>_{\kz^{2E}} \\
&= \left<D_{(\ba\bb)}^{\frac{1}{2}}\Phi_{\mathrm{bv}},\ui I_{\pm}D_{(\ba\bb)}^{\frac{1}{2}}\Psi_{\mathrm{bv}}\right>_{\kz^{2E}} \\
&= \left<UD_{(\ba\bb)}^{\frac{1}{2}}\Phi_{\mathrm{bv}},JUD_{(\ba\bb)}^{\frac{1}{2}}\Psi_{\mathrm{bv}}\right>_{\kz^{2E}} \quad \mbox{for all} \quad \psi,\phi\in H^1(\Gamma).
}
\end{equation}
To define the square root of $D_{(\ba\bb)}$ we have used the usual definition of a positive operator (\cite[p.\,196]{Reed:1980}), which in this case simply means to take the square root of the (diagonal) entries in $D_{(\ba\bb)}$. Note that 
\begin{equation}
\label{45}
\omega(\cdot,\cdot)\coloneqq 
\left<\cdot,J\cdot\right>_{\kz^{2E}}
\end{equation}
defines a nondegenerate complex symplectic form on $\kz^{2E}\times\kz^{2E}$. We call a subspace $\boldsymbol{\mathcal{L}}$ of $\kz^{2E}$ a Lagrangian subspace iff 
\begin{equation}
\label{17}
\eqalign{
\bullet \ \ & a,b\in\boldsymbol{\mathcal{L}} \quad \mbox{then} \quad \omega(a,b)=0. \\
\bullet \ \ & \mbox{Whenever for a subspace} \ \tilde{\boldsymbol{\mathcal{L}}}\supset\boldsymbol{\mathcal{L}} \ \mbox{the first property holds,}\\
& \mbox{it follows} \ \tilde{\boldsymbol{\mathcal{L}}}=\boldsymbol{\mathcal{L}}.
}
\end{equation}
For the Lagrangian subspaces of $\kz^{2E}$ we apply the result of \cite{KostrykinSchrader:1999}. A subspace $\boldsymbol{\mathcal{L}}$ is Lagrangian iff there exist two matrices $A,B\in\Mat(E\times E,\kz)$ with:
\begin{equation}
\label{18}
\eqalign{
& AB^+=BA^+ \quad \mbox{and} \\
& \rank(A,B)=E.
}
\end{equation}
We then have,
\begin{equation}
\label{85}
\boldsymbol{\mathcal{L}}=\left\{\phi\in\kz^{2E}; \ \phi\coloneqq 
\left(
\begin{array}{c}
\phi_1 \\
\phi_2
\end{array}
\right)
\quad \mbox{and} \quad A\phi_1+B\phi_2=0\right\}.
\end{equation}
In (\ref{18}) the matrix $(A,B)$ is formed of the columns of $A$ and $B$, and we have introduced two maps 
\begin{equation}
\label{2020}
(\boldsymbol{\cdot})_i :\kz^{2E}\rightarrow\kz^{E} \quad \mbox{for} \quad 1\leq i\leq 2
\end{equation}
by
\begin{equation}
\label{40}
\phi_i\coloneqq 
\cases{
(\eins,0)\phi & if $i=1$, \\
(0,\eins)\phi & if $i=2$.
}
\end{equation}
Furthermore, as mentioned in \cite{KostrykinSchrader:2006b}, these matrices are not uniquely defined. Two sets of matrices $A,B$ and $\widetilde{A},\widetilde{B}$ define the same Lagrangian subspace iff there exists an invertible matrix $C$ with 
\begin{equation}
\label{2021}
A=C\widetilde{A} \quad \mbox{and} \quad B=C\widetilde{B}.
\end{equation}
Since $UD_{(\ba\bb)}^{\frac{1}{2}}$ is a bijection from $\kz^{2E}$ onto itself and by (\ref{13}) and (\ref{19}) we infer, with the same arguments as in \cite{KostrykinSchrader:1999}, that there is a one-to-one correspondence between the self-adjoint extensions of $\left(H_{\mathrm{BK}},D_{0}^1(\Gamma)\right)$ and the Lagrangian subspaces of $\kz^{2E}$. We thus have proved the following proposition.
\begin{prop}
\label{ab1}
Each domain of definition of such a self-adjoint extension is exactly the preimage with respect to (\ref{15}) of a subspace 
\begin{equation}
\label{20}
\boldsymbol{\mathcal{L}}=\left\{\Psi_{\mathrm{bv}}\in\kz^{2E}; \ A\left(UD_{(\ba\bb)}^{\frac{1}{2}}\Psi_{\mathrm{bv}}\right)_1+B\left(UD_{(\ba\bb)}^{\frac{1}{2}}\Psi_{\mathrm{bv}}\right)_2=0\right\},
\end{equation}
where $A$ and $B$ fulfil (\ref{18}). The converse is also true.
\end{prop}
Because of proposition \ref{ab1}, we denote in the following the self-adjoint extensions of $\left(H_{\mathrm{BK}},D_{0}^1(\Gamma)\right)$ by $\left(H_{\mathrm{BK}};A,B\right)$.
\section{Determination of the eigenvalues of $H_{\mathrm{BK}}$}
\label{66}
All possible eigenfunctions $\psi_{k}$ to an eigenvalue $k$ of $H_{\mathrm{BK}}$ are of the form
\begin{equation}
\label{2022}
\psi_k(x)=\left(\alpha_1\frac{1}{\sqrt{x}}\ue^{\ui k\ln x},\ldots,\alpha_{E}\frac{1}{\sqrt{x}}\ue^{\ui k\ln x}\right).
\end{equation}
We denote the column vector (\ref{15}) corresponding to $\psi_k$ by $\Psi_{bv,k}$. In order to apply (\ref{20}) for determining the eigenvalues $k$ and the corresponding eigenvectors $\psi_k$, we calculate $UD_{(\ba\bb)}^{\frac{1}{2}}\Psi_{bv,k}$ using (\ref{21}).
\begin{equation}
\label{3300}
\eqalign{
UD_{(\ba\bb)}^{\frac{1}{2}}\Psi_{bv,k} &= U
\left(
\begin{array}{cc}
\ue^{\ui k\ln \ba} & 0 \\
0 & \ue^{\ui k \ln \bb}
\end{array}
\right)
\ap \\
&=\frac{1}{\sqrt{2}} 
\left(
\begin{array}{cc}
\ui\ue^{\ui k\ln \ba} & \ue^{\ui k \ln \bb} \\
-\ue^{\ui k \ln \ba} & -\ui\ue^{\ui k \ln \bb}
\end{array}
\right)
\ap \\
&=\frac{1}{\sqrt{2}}  
\left(
\begin{array}{cc}
\ui\ue^{\ui k\ln \ba}+ \ue^{\ui k \ln \bb}& 0 \\
0 & -\ue^{\ui k \ln \ba}-\ui\ue^{\ui k \ln \bb}
\end{array}
\right)
\ap.
}
\end{equation}
For convenience, we have used the notations:
\begin{equation}
\label{24}
\eqalign{
\left(\ue^{{\ui k\ln \ba}}\right)_{mn} & \coloneqq \delta_{mn}\ue^{{\ui k\ln a_m}},\\
\left(\ue^{{\ui k\ln \bb}}\right)_{mn} & \coloneqq \delta_{mn}\ue^{{\ui k\ln b_m}} \quad \mbox{for}  \quad 1\leq m,n\leq E \quad 
}
\end{equation}
and
\begin{equation}
\label{2023}
\ap\coloneqq (\alpha_1,\ldots,\alpha_E,\alpha_1,\ldots,\alpha_E)^T.
\end{equation}
Therefore, we get:
\begin{equation}
\label{22}
\eqalign{
\sqrt{2}\left(UD_{(\ba\bb)}^{\frac{1}{2}}\Psi_{bv,k}\right)_1 &=\left(\ui\ue^{{\ui k\ln \ba}}+\ue^{{\ui k\ln \bb}}\right)\ap_1 \\
\sqrt{2}\left(UD_{(\ba\bb)}^{\frac{1}{2}}\Psi_{bv,k}\right)_2 &=-\left(\ue^{{\ui k\ln \ba}}+\ui\ue^{{\ui k\ln \bb}}\right)\ap_2,
}
\end{equation}
where we have used $(\boldsymbol{\cdot})_i$ defined in (\ref{40}).

Taking into account that $\ap_1=\ap_2=: \tilde{\ap}$, we obtain for the expression in (\ref{20})
\begin{equation}
\label{3301}
\eqalign{
& \sqrt{2}\left[A\left(UD_{(\ba\bb)}^{\frac{1}{2}}\Psi_{bv,k}\right)_1+B\left(UD_{(\ba\bb)}^{\frac{1}{2}}\Psi_{bv,k}\right)_2\right] \\
& \hspace{30mm} =\Bigl(A\left(\ui\ue^{{\ui k\ln \ba}}+\ue^{{\ui k\ln \bb}}\right)-B\left(\ue^{{\ui k\ln \ba}}+\ui\ue^{{\ui k\ln \bb}}\right)\Bigr)\tilde{\ap} \\
& \hspace{30mm} =\left(\ui\left(A+\ui B\right)\ue^{{\ui k\ln \ba}}+\left(A-\ui B\right)\ue^{{\ui k\ln \bb}}\right)\tilde{\ap}.
}
\end{equation}
Kostrykin and Schrader have shown, see \cite{KostrykinSchrader:1999}, that $A\pm \ui B$ are invertible under the assumption (\ref{18}). With the notation
\begin{equation}
\label{2025}
\boldsymbol{C}(k)\coloneqq \ue^{{\ui k\ln \ba}}\tilde{\ap}
\end{equation}
and, because $(A-\ui B)$ and $(A+\ui B)^{-1}$ commute (see \cite{KostrykinSchrader:2006}), we get:
\begin{equation}
\label{23}
\eqalign{
 & A\left(UD_{(\ba\bb)}^{\frac{1}{2}}\Psi_{bv,k}\right)_1+B\left(UD_{(\ba\bb)}^{\frac{1}{2}}\Psi_{bv,k}\right)_2=0 \\
\Leftrightarrow & \left(\eins-\ui\frac{A-\ui B}{A+\ui B}\ue^{{\ui k\ln \frac{\bb}{\ba}}}\right)\boldsymbol{C}(k)=0,
}
\end{equation}
with $\ue^{{\ui k\ln \frac{\bb}{\ba}}}$ similarly defined as in (\ref{24}). 
Due to the similarity of (\ref{23}) with the secular equation for the common Laplace operator on compact graphs (see \cite{KostrykinSchrader:2006b}), we denote:
\begin{equation}
\label{25}
\mathcal{S}(A,B)\coloneqq \ui\frac{A-\ui B}{A+\ui B} \quad \mbox{and} \quad \mathcal{T}\left(\ba,\bb;k\right)\coloneqq \ue^{{\ui k\ln \frac{\bb}{\ba}}}.
\end{equation}
It follows with exactly the same arguments as in \cite{KostrykinSchrader:1999} that $\mathcal{S}(A,B)$ is unitary, and we shall call it also the $S$-matrix of the quantum graph. The unitarity of $\mathcal{T}\left(\ba,\bb;k\right)$ iff $k\in\rz$ is obvious. Since, for all $k\in\kz$, $\boldsymbol{C}(k)=0$ iff $\tilde{\ap}=0$, we have proved the following proposition establishing the secular equation $\mathcal{F}(k)=0$.
\begin{prop}
\label{ab2}
$k_n\in\rz$ is an eigenvalue of $\left(H_{\mathrm{BK}};A,B\right)$ iff
\begin{equation} 
\label{75}
\mathcal{F}(k_n)\coloneqq \det\bigl(\eins_{E\times E}-\mathcal{S}(A,B)\mathcal{T}\left(\ba,\bb;k_n\right)\bigr)=0.
\end{equation}
Furthermore, the multiplicity of the eigenvalue one of $\mathcal{S}(A,B)\mathcal{T}\left(\ba,\bb;k_n\right)$ coincides with the multiplicity of the eigenvalue $k_n$ of $\left(H_{\mathrm{BK}};A,B\right)$.
\end{prop}
We remark that the geometric multiplicities and the algebraic multiplicities of $\mathcal{S}(A,B)\mathcal{T}\left(\ba,\bb;k\right)$ coincide since this matrix is diagonalizable.

In contrast to the $S$-matrix of the generic negative Laplacian $-\Delta$ on graphs, the S-matrix $\mathcal{S}(A,B)$ is always independent of $k$ (the $S$-matrix of $-\Delta$ is independent of the wave number $k$ iff $S^+=S$ [see \cite{KostrykinSchrader:2006b}]). But we remark that the independence of the $S$-matrix on the eigenvalue will also occur when we replace $H_{\mathrm{BK}}$ by the standard momentum operator (with $x\in\rz$)
\begin{equation}
\label{2030}
p\coloneqq -\ui\frac{\ud\phantom{x}}{\ud x}.
\end{equation}
The calculations are quite analogous (see e.g. \cite{Carlson:1999} for a detailed discussion of this). In fact every self-adjoint extension of $p$ can be characterized by the same matrices $A$ and $B$ as in (\ref{18}) and we would get the same $S$-matrix $\mathcal{S}(A,B)$. The only difference in the secular equation between the operators $H_{\mathrm{BK}}$ and $p$ then is the form of the second matrix in (\ref{25}) which in the case of $p$ is given by
\begin{equation}
\label{2026}
T(\ba,\bb;k)\coloneqq \ue^{{\ui k(\bb-\ba)}}.
\end{equation}
This is one reason why we rather relate $H_{\mathrm{BK}}$ with a momentum operator than an energy operator as indicated in section \ref{26}. Because of the occurrence of the logarithm in $\mathcal{T}$ we endow the quantum graph with a metric structure as proposed in section \ref{7}. We also mention that proposition \ref{ab2} is valid for all $k\in\kz$, in particular for $k=0$ in contrast to the corresponding proposition \ref{ab4} for $H_{BK}^2$. However, the analogy between $p^2=-\Delta$ and $H_{\mathrm{BK}}^2$ is not so obvious.
\section{The ``squared'' Berry-Keating operator}
\label{27}
Our Hilbert space will be $\mathcal{H}=L^2(\Gamma)$, see (\ref{32}) and (\ref{33}). Again, we seek self-adjoint extensions of (\ref{30}) with respect to $C_0^{\infty}(\Gamma)$. In order to obtain a self-adjoint operator, the task is to specify an appropriate domain $D(H_{\mathrm{BK}}^2)$ for this operator with 
\begin{equation}
\label{3031}
C_0^{\infty}(\Gamma)\subset D(H_{\mathrm{BK}}^2).
\end{equation}
One simple possibility is to define $H_{\mathrm{BK}}^2$ as the ``squared'' Berry-Keating operator, which means:
\begin{equation}
\label{36}
\eqalign{
H_{\mathrm{BK}}^2\psi & \coloneqq H_{\mathrm{BK}}\left(H_{\mathrm{BK}}\psi\right),\\
\psi\in D(H_{\mathrm{BK}}^2) & \coloneqq \bigl\{\phi\in D(H_{\mathrm{BK}}); \ H_{\mathrm{BK}}\phi\in D(H_{\mathrm{BK}}) \bigr\}. 
}
\end{equation}
It follows immediately that $H_{\mathrm{BK}}^2$ is self-adjoint if $H_{\mathrm{BK}}$ is self-adjoint using Friedrichs' extension theorem \cite[p.\,180]{Reed:1975}. But in fact there are many possible self-adjoint extensions which cannot be realized in such a way. We will give simple examples in section \ref{120}. We can generalize these constructions to consider non-self-adjoint but closed realizations of $H_{\mathrm{BK}}$ and then form 
\begin{equation}
\label{3005}
H_{\mathrm{BK}}^+H_{\mathrm{BK}} \quad \mbox{or} \quad H_{\mathrm{BK}}H_{\mathrm{BK}}^+.
\end{equation}
This is an idea quite analogous to the concept of supersymmetry, see \cite{Thaller:1992} and \cite{Fulling:2007} (the technique of factorization was already introduced by Schrödinger \cite{Schrödinger:1940} and reviewed in \cite{Infeld:1951}). In \cite{Fulling:2007} this technique has been used but isn't explicitly mentioned. However, only a certain kind of self-adjoint extension can be attained in such a way. In \cite{Fulling:2007} these are exactly the self-adjoint extensions which correspond to $k$-independent $S$-matrices corresponding to these extensions. This relation between the $S$-matrices and the self-adjoint extensions of the negative Laplace operator $-\Delta$ on metric graphs is explained in \cite{KostrykinSchrader:2006b}. 

We would like to give an overview of the starting point of our considerations from a mathematical point of view. The proofs of these statements are similar as in section \ref{9} using the same references as there. Therefore, we only summarize the results. First, the operator $H_{\mathrm{BK}}^2$ acting on $C_0^{\infty}(\Gamma)$ or 
\begin{equation}
\label{37}
D_0^2(\Gamma)\coloneqq \bigoplus\limits_{i=1}^{E}H_0^2[a_i,b_i] \quad \mbox{with} \quad 0<a_i<b_i<\infty
\end{equation}
is symmetric. $H_0^2[a_i,b_i]$ is the set of absolutely continuous functions which possess absolutely continuous derivatives on $[a_i,b_i]$, square integrable weak second derivatives and which together with their first derivatives vanish at the endpoints of the intervals. Furthermore, $H_{\mathrm{BK}}^2$ acting on $D_0^2(\Gamma)$ is closed and the adjoint operator of $\left(H_{\mathrm{BK}}^2,D_0^2(\Gamma)\right)$ is $\left(H_{\mathrm{BK}}^2,H^2(\Gamma)\right)$. Here
\begin{equation}
\label{42}
H^2(\Gamma)\coloneqq \bigoplus\limits_{i=1}^E H^2[a_i,b_i] \quad \mbox{with} \quad 0<a_i<b_i<\infty
\end{equation}
is the space of functions being absolutely continuous on the intervals of the graph $\Gamma$ possessing absolutely continuous derivatives and weak square integrable second derivatives. The deficiency indices are $(2E,2E)$, thus $(H_{\mathrm{BK}}^2,D_0^2(\Gamma))$ possesses infinitely many self-adjoint extensions. Again, as for $H_{BK}$ the spectrum of every self-adjoint extension is purely discrete and as in section \ref{9} the projections of the spaces $D_0^2(\Gamma)$ and $H^2(\Gamma)$ on the intervals of the graph $\Gamma$ coincide with the corresponding Sobolev spaces (see again e.g.\ \cite{Reed:1980}). We shall follow a general approach to find all these self-adjoint extensions, quite analogous as in section \ref{14} and based on \cite{KostrykinSchrader:1999}.
\section{Classification of the self-adjoint extensions of the ``squared'' Berry-Keating operator}
\label{38}
First, we define $\Psi_{\mathrm{bv}}$ as in (\ref{15}) for $\psi\in H^2(\Gamma)$ and additionally 
\begin{equation}
\label{39}
\eqalign{
\Psi_{\mathrm{bv}}'& \coloneqq \bigl(\psi_1'(a_1),\ldots,\psi_E'(a_E),-\psi_{1}'(b_1),\ldots,-\psi_{E}'(b_{E})\bigr)^T\\
& \mbox{for} \quad \psi\in H^2(\Gamma),
}
\end{equation}
in which $\psi_i'$ is the derivative of $\psi_i$ on the interval $I_i$. Similarly as in (\ref{13}), we define a symplectic form on $H^2(\Gamma)\times H^2(\Gamma)$
\begin{equation}
\label{59}
\left[\phi,\psi\right]_2\coloneqq \left<\phi,{H_{\mathrm{BK}}^2}^+\psi\right>_{L^2(\Gamma)}-\left<{H_{\mathrm{BK}}^2}^+\phi,\psi\right>_{L^2(\Gamma)} \quad  \mbox{for} \ \phi,\psi\in H^2(\Gamma).
\end{equation}
With the same arguments as for $H_{\mathrm{BK}}$ in section \ref{9} the task is to find all maximal $[\cdot,\cdot]_2$-symmetric subspaces of $H^2(\Gamma)$ in order to find all self-adjoint extensions to $(H_{\mathrm{BK}}^2,D_0^2(\Gamma))$. We shall adapt the definition of $J$ in (\ref{21}) by 
\begin{equation}
\label{3010}
J\coloneqq 
\left(
\begin{array}{cc}
0 & \eins_{2E\times 2E} \\
-\eins_{2E\times 2E} & 0
\end{array}
\right)
\end{equation}
and (see (\ref{16}))
\begin{equation}
\label{43}
\tD\coloneqq 
\left(
\begin{array}{cc}
\D & 0 \\
0 & \D
\end{array}
\right), \quad
\Ps_{\mathrm{bv}}\coloneqq 
\left(
\begin{array}{cc}
\Psi_{\mathrm{bv}} \\
\Psi_{\mathrm{bv}}'
\end{array}
\right)
\quad \mbox{for} \quad \psi\in D_0^2(\Gamma).
\end{equation}
We obtain for $\phi,\psi\in H^2(\Gamma)$ using partial integration and the fact that $J$ and $\tD$ commute:
\begin{equation}
\label{44}
\fl\eqalign{
\left[\psi,\phi\right]_2 &= \sum\limits_{i=1}^{E}\biggl(b_i^2\Bigl(\overline{\psi_i'(b_i)}\phi(b_i)-\overline{\psi_i(b_i)}\phi'(b_i)\Bigr)-a_i^2\Bigl(\overline{\psi_i'(a_i)}\phi(a_i)-\overline{\psi_i(a_i)}\phi'(a_i)\Bigr)\biggr) \\
&= \left<[\psi]_{\mathrm{bv}},J \tD^2[\phi]_{\mathrm{bv}}\right>_{\kz^{4E}} \\
&= \left<\tD[\psi]_{\mathrm{bv}},J \tD[\phi]_{\mathrm{bv}}\right>_{\kz^{4E}}.
}
\end{equation}
Taking the scalar product in the definition of $\omega(\cdot,\cdot)$ in (\ref{45}) with respect to $\kz^{4E}$, we infer as in section \ref{14} the following proposition.
\begin{prop}
\label{ab3}
The self-adjoint extensions of $(H_{\mathrm{BK}}^2,D_0^2(\Gamma))$ are exactly the preimages of 
\begin{equation}
\label{46}
\eqalign{\boldsymbol{\mathcal{L}} &=\left\{[\phi]_{\mathrm{bv}}\in\kz^{4E}; \ A\left(\tD[\phi]_{\mathrm{bv}}\right)_1+B\left(\tD[\phi]_{\mathrm{bv}}\right)_2=0\right\} \\
                         &=\left\{[\phi]_{\mathrm{bv}}\in\kz^{4E}; \ A\D\Phi_{\mathrm{bv}}+B\D\Phi_{\mathrm{bv}}'=0\right\}
}
\end{equation}
with respect to $[\boldsymbol{\cdot}]_{\mathrm{bv}}$ in (\ref{43}).
\end{prop}
In (\ref{46}) we have used $(\boldsymbol{\cdot})_i$ defined in (\ref{40}). The matrices $A$ and $B$ are now elements of $\Mat(2E\times 2E,\kz)$ with the adopted conditions
\begin{equation}
\label{47}
\eqalign{& AB^+=BA^+ \quad \mbox{and} \\
& \rank(A,B)=2E.
}
\end{equation}
Again, as in section \ref{14} and because of proposition \ref{ab3} we denote the self-adjoint extensions of $\left(H_{\mathrm{BK}}^2,D_{0}^2(\Gamma)\right)$ by $\left(H_{\mathrm{BK}}^2;A,B\right)$. 
\section{Determination of the eigenvalues of $H_{\mathrm{BK}}^2$}
We want to solve the eigenvalue problem
\begin{equation}
\label{50}
H_{\mathrm{BK}}^2\psi=\lambda\psi.
\end{equation}
To tackle this problem, it will be convenient to consider the wave number $k$ defined by $\lambda^{\frac{1}{2}}\coloneqq k$. It is a trivial observation that $\pm k$ correspond to the same eigenvalue $\lambda$. This fact will be revealed in the symmetry of the secular equation for the wave number. 

In addition to (\ref{50}) the eigenvector $\psi$ must be in the domain of definition of the operator. However, the general form of the eigenvector to an eigenvalue $\lambda=k^2\neq 0$ is 
\begin{equation}
\label{51}
\psi_k(x)=\left(\frac{1}{\sqrt{x}}\left(\alpha_1\ue^{\ui k\ln x}+\beta_1\ue^{-\ui k\ln x}\right),\ldots,\frac{1}{\sqrt{x}}\left(\alpha_E\ue^{\ui k\ln x}+\beta_E\ue^{-\ui k\ln x}\right)\right).
\end{equation}
We can proceed as in \cite{KostrykinSchrader:1999}. Therefore, we compute $\Psi_{bv,k}$ and $\Psi_{bv,k}'$ using the definitions in (\ref{16}) and (\ref{24}). 
\begin{equation}
\label{52}
\eqalign{
\Psi_{bv,k} &= \D^{-\frac{1}{2}}
\left(
\begin{array}{cc}
\ue^{\ui k \ln \ba} & \ue^{-\ui k \ln \ba} \\
\ue^{\ui k \ln \bb} & \ue^{-\ui k \ln \bb}
\end{array}
\right)
\left(
\begin{array}{cc}
\ap \\
\bp
\end{array}
\right)
\\
\Psi_{bv,k}' &= \left(-\frac{1}{2}\D^{-\frac{3}{2}}
\left(
\begin{array}{cc}
\ue^{\ui k \ln \ba} & \ue^{-\ui k \ln \ba} \\
-\ue^{\ui k \ln \bb} & -\ue^{-\ui k \ln \bb}
\end{array}
\right)
\right.\\
&\left. \hspace{3cm}+\ui k \D^{-\frac{3}{2}}
\left(
\begin{array}{cc}
\ue^{\ui k \ln \ba} & \ue^{-\ui k \ln \ba} \\
-\ue^{\ui k \ln \bb} & -\ue^{-\ui k \ln \bb}
\end{array}
\right)
I_{\pm}\right)
\left(
\begin{array}{cc}
\ap \\
\bp
\end{array}
\right).
}
\end{equation}
In order to be in the domain of definition of a self-adjoint realization, $\Psi_{bv,k}$ and $\Psi_{bv,k}'$ must be in some $\boldsymbol{\mathcal{L}}$ of (\ref{46}) defined by the two matrices $A$ and $B$. We make the identification (cf.\ \cite{KostrykinSchrader:2006b}),
\begin{equation}
\label{3011}
X(k;\ba,\bb)\coloneqq 
\left(
\begin{array}{cc}
\ue^{\ui k \ln \ba} & \ue^{-\ui k \ln \ba} \\
\ue^{\ui k \ln \bb} & \ue^{-\ui k \ln \bb}
\end{array}
\right),
\end{equation}
\begin{equation}
\label{3012}
Y(k;\ba,\bb)\coloneqq 
\left(
\begin{array}{cc}
\ue^{\ui k \ln \ba} & \ue^{-\ui k \ln \ba} \\
-\ue^{\ui k \ln \bb} & -\ue^{-\ui k \ln \bb}
\end{array}
\right)
\quad \mbox{and} 
\end{equation}
\begin{equation}
Y'(k;\ba,\bb)\coloneqq 
\left(
\begin{array}{cc}
\ue^{\ui k \ln \ba} & -\ue^{-\ui k \ln \ba} \\
-\ue^{\ui k \ln \bb} & \ue^{-\ui k \ln \bb}
\end{array}
\right).
\end{equation}
Thus, we conclude, with the definition for the bold symbols in accordance with (\ref{60}):
\begin{equation}
\label{62}
0\stackrel{!}{=}A\D\Phi_{\mathrm{bv}} + B\D\Phi_{\mathrm{bv}}' 
\end{equation}
\begin{equation}
\label{62a}
\fl\eqalign{
=& \left(A\D^{\frac{1}{2}}X(k;\ba,\bb)+B\D^{-\frac{1}{2}}Y(k;\ba,\bb)
\left(
\begin{array}{cc}
\boldsymbol{-\frac{1}{2}+\ui k} & 0 \\
0 & \boldsymbol{-\frac{1}{2}-\ui k}
\end{array}
\right)
\right)
\left(
\begin{array}{cc}
\ap \\
\bp
\end{array}
\right)
\\
=& \left(\Bigl(A\D^{\frac{1}{2}}-\frac{1}{2}B\D^{-\frac{1}{2}}I_{\pm}\Bigr)X(k;\ba,\bb) +\ui kB\D^{-\frac{1}{2}}Y'(k;\ba,\bb)\right)
\left(
\begin{array}{cc}
\ap \\
\bp
\end{array}
\right).
}
\end{equation}

At this point we make two observations: Since $\D^{-\frac{1}{2}}$ is self-adjoint, we conclude
\begin{equation}
\label{58}
A\D^{\frac{1}{2}}\left(B\D^{-\frac{1}{2}}\right)^+=AB^+.
\end{equation}
Since $\D^{-\frac{1}{2}}$ and $\D^{\frac{1}{2}}$ are invertible and diagonal, it is easy to show that
\begin{equation}
\label{56}
\rank(A\D^{\frac{1}{2}},B\D^{-\frac{1}{2}})=\rank(A,B)=2E.
\end{equation}
Therefore, we define
\begin{equation}
\label{57}
A\D^{\frac{1}{2}}=:A' \quad \mbox{and} \quad B\D^{-\frac{1}{2}}=: B'
\end{equation}
and observe that $A'$ and $B'$ also fulfil the conditions (\ref{47}). Therefore, we can apply a theorem of Kuchment \cite{Kuchment:2004}. It states that two matrices $A'$ and $B'$ fulfil (\ref{47}) iff there exists an invertible matrix $C$ with:
\begin{equation}
\label{61}
A'=C\PBb+ C\PBs L'\PBs \quad \mbox{and} \quad B'=C\PBs.
\end{equation}
In (\ref{61}) we have defined $\PBb$ as the projector onto the kernel of $B'$ and $\PBs$ as the corresponding orthogonal projector. The matrix $L'$ is self-adjoint and can be defined by (see \cite{Kuchment:2004}), 
\begin{equation}
\label{3020}
L'\coloneqq \left(\left.B'\right|_{\ran B'^+}\right)^{-1}A'\PBs.
\end{equation}
Hence, we can proceed in the calculation (\ref{62}) by multiplying (\ref{62a}) from the left-hand side by $C^{-1}$
\begin{equation}
\label{64}
\eqalign{
0 &=\left(\Bigl(\PBb+ \PBs L'\PBs-\frac{1}{2}\PBs I_{\pm}\Bigr)X(k;\ba,\bb)\right.\\
& \hspace{3cm}\left.\phantom{\frac{1}{2}\PBs}+\ui k\PBs Y'(k;\ba,\bb)\right)
\left(
\begin{array}{c}
\ap \\
\bp
\end{array}
\right).
}
\end{equation}
Since the projectors $\PBb$ and $\PBs$ are mutually orthogonal, we infer from (\ref{61}), the definition of $\boldsymbol{\mathcal{L}}$ in (\ref{46}) and with the first line of (\ref{62a}), that
\begin{equation}
\label{63}
\PBb
X(k;\ba,\bb)
\left(
\begin{array}{c}
\ap \\
\bp
\end{array}
\right)
=0.
\end{equation}
In (\ref{64}) we insert between the matrices $I_{\pm}$ and $X(k;\ba,\bb)$ the unit matrix $\eins=\PBb+\PBs$ and apply (\ref{63})
\begin{equation}
\label{65}
\eqalign{
0 & =\left(\Bigl(\PBb+ \PBs \Bigl(L'-\frac{1}{2}I_{\pm}\Bigr)\PBs\Bigr)X(k;\ba,\bb)\right.\\
& \left.\hspace{3cm}\phantom{-\frac{1}{2}I_{\pm}} +\ui k\PBs Y'(k;\ba,\bb)\right)
\left(
\begin{array}{c}
\ap \\
\bp
\end{array}
\right).
}
\end{equation}
We realize that $L'-\frac{1}{2}I_{\pm}$ is also self-adjoint. Thus, we define
\begin{equation}
\label{72}
L''\coloneqq \PBs\left(L'-\frac{1}{2}I_{\pm}\right)\PBs \quad 
\end{equation}
and make a re-definition:
\begin{equation}
\label{71}
A''\coloneqq \PBb+ L'' \quad \mbox{and} \quad B''\coloneqq \PBs.
\end{equation}
It is obvious that the matrices $A''$ and $B''$ fulfil the conditions (\ref{44}). Hence as in section \ref{66} respectively \cite{KostrykinSchrader:1999}, we infer that $A''\pm\ui k B''$ is invertible and conclude with quite the same calculation as in \cite{KostrykinSchrader:2006b}
\begin{equation}
\label{83}
\fl0=\left(A''+\ui kB'' \right)\left[\eins-S''(A,B;k)T(\ba,\bb;k)\right]
\left(
\begin{array}{cc}
\ue^{\ui k \ln \ba} & 0 \\
0 & \ue^{-\ui k \ln \bb}
\end{array}
\right)
\left(
\begin{array}{c}
\ap \\
\bp
\end{array}
\right).
\end{equation}
Here we have used the definitions
\begin{equation}
\label{67}
\eqalign{
S''(A,B;k) & \coloneqq S(A'',B'';k)\coloneqq -\frac{A''-\ui kB''}{A''+\ui kB''} \quad \mbox{and}\\ 
T(\ba,\bb;k) & \coloneqq 
\left(
\begin{array}{cc}
0 & \ue^{{\ui k\ln \frac{\bb}{\ba}}} \\
\ue^{{\ui k\ln \frac{\bb}{\ba}}} & 0
\end{array}
\right)
.
}
\end{equation}

The first and the third matrix in the product of (\ref{83}) are invertible for all $k\in \kz\setminus\left(\pm\ui\sigma(L'')\right)$ in which $\sigma(L'')$ denotes the spectrum of $L''$. For a detailed discussion of this, see \cite{BE:2008} and \cite{KostrykinSchrader:2006}. Thus, we have proved the following proposition.
\begin{prop}
\label{ab4}
$k^2$ with $k\in\kz\setminus\left(\pm\ui\sigma(L'')\cup\{0\}\right)$ is an eigenvalue of $\left(H_{\mathrm{BK}}^2;A,B\right)$ iff
\begin{equation}
\label{68}
F(k):=\det\left(\eins_{2E\times2E}-S''(A,B;k)T(\ba,\bb;k)\right)=0.
\end{equation}
Furthermore, as in section \ref{66}, the multiplicity of the eigenvalue $\lambda=k^2$ coincides with the multiplicity of the eigenvalue one of $S''(A,B;k)T(\ba,\bb;k)$ for every $k\in\kz\setminus\left(\pm\ui\sigma(L'')\cup\{0\}\right)$.
\end{prop}
We remark that the restriction on $k$ concerns only non-positive eigenvalues $\lambda=k^2$ of $H_{\mathrm{BK}}^2$.  
\section{The eigenvalue zero}
\label{90}
For the eigenvalue $\lambda=0$, which is equivalent to the case $k=0$, the eigenfunctions are of the form
\begin{equation}
\label{3021}
\psi_0(x)=\left(\alpha_1\frac{1}{\sqrt{x}}+\beta_1\frac{1}{\sqrt{x}}\ln x,\ldots,\alpha_E\frac{1}{\sqrt{x}}+\beta_E\frac{1}{\sqrt{x}}\ln x\right).
\end{equation}
With a similar calculation as for the case $k\neq0$ one obtains the equation
\begin{equation}
\label{3022}
\left(A''
\left(
\begin{array}{cc}
\eins_{E\times E} & \ln \ba \\
\eins_{E\times E} & \ln \bb
\end{array}
\right)
+B''
\left(
\begin{array}{cc}
0 & \eins_{E\times E} \\
0 & -\eins_{E\times E}
\end{array}
\right)
\right)
\left(
\begin{array}{cc}
\ap \\
\bp
\end{array}
\right)
=0
\end{equation}
which is necessary and sufficient for $\lambda=0$ to be an eigenvalue of $\left(H_{\mathrm{BK}}^2;A,B\right)$. The matrices $A''$ and $B''$ are the same as in (\ref{71}). Then we can proceed as in \cite{BE:2008} and get the following proposition.
\begin{prop}
\label{ab5}
$\lambda=k^2=0$ is an eigenvalue of $\left(H_{\mathrm{BK}}^2;A,B\right)$ iff for one value $k'\neq0$ and then for every $k'\neq 0$ 
\begin{equation}
\label{91}
F_0(k')\coloneqq \det\bigl(\eins-S''(A,B;k')C(\ba,\bb;k')\bigr)=0
\end{equation}
is fulfilled with $S''(A,B;k')$ as in (\ref{67}) and 
\begin{equation}
\label{3030}
\fl\eqalign{
C(\ba,\bb;k') & \coloneqq \\
& \left(
\begin{array}{cc}
\begin{array}{ccc}
\frac{\ln\left(\frac{b_1}{a_1}\right)}{2\frac{\ui}{k'}+\ln\left(\frac{b_1}{a_1}\right)} & & 0\\
 & \ddots & \\
 0 & & \frac{\ln\left(\frac{b_E}{a_E}\right)}{2\frac{\ui}{k'}+\ln\left(\frac{b_E}{a_E}\right)}
\end{array}
&
\begin{array}{ccc}
\frac{2\frac{\ui}{k'}}{2\frac{\ui}{k'}+\ln\left(\frac{b_1}{a_1}\right)} & & 0\\
 & \ddots & \\
 0 & & \frac{2\frac{\ui}{k'}}{2\frac{\ui}{k'}+\ln\left(\frac{b_E}{a_E}\right)}
\end{array}
\\
\begin{array}{ccc}
\frac{2\frac{\ui}{k'}}{2\frac{\ui}{k'}+\ln\left(\frac{b_1}{a_1}\right)} & &0 \\
 & \ddots & \\
 0& & \frac{2\frac{\ui}{k'}}{2\frac{\ui}{k'}+\ln\left(\frac{b_E}{a_E}\right)}
\end{array}
&
\begin{array}{ccc}
\frac{\ln\left(\frac{b_1}{a_1}\right)}{2\frac{\ui}{k'}+\ln\left(\frac{b_1}{a_1}\right)} & & 0\\
 & \ddots & \\
 0 & & \frac{\ln\left(\frac{b_E}{a_E}\right)}{2\frac{\ui}{k'}+\ln\left(\frac{b_E}{a_E}\right)}
\end{array}
\end{array}
\right).
}
\end{equation}
Furthermore, the multiplicity of the eigenvalue $\lambda=0$ coincides with the multiplicity of the eigenvalue one of $S''(A,B;k')C(\ba,\bb;k')$ for every real $k'\neq0$.
\end{prop}
Thus, in general there is a difference between the spectral multiplicity of the eigenvalue one of $S''(A,B;0)T(\ba,\bb,0)$, which we denote by $N$, and the eigenvalue one of $S''(A,B;k')C(\ba,\bb;k')$ with $k'\neq0$, see \cite{BE:2008}, \cite{Fulling:2007} and \cite{Kurasov:2005}.

In order to relate the self-adjoint extensions of $(H_{\mathrm{BK}}^2,D_{0}^2(\Gamma))$ with the self-adjoint extensions of the Laplacian $-\Delta$ one has to adjust the lengths as before. However, in oder to attain the same spectrum, except for the case $k\notin\kz\setminus\left(\pm\ui\sigma(L'')\setminus\{0\}\right)$, one has to transform the matrices $A$ and $B$ into $A''$ and $B''$ as in (\ref{57}) and (\ref{71}). Then the spectrum of the negative Laplacian characterized by $A''$ and $B''$ with the previous choice of the lengths will coincide with the spectrum of $H_{\mathrm{BK}}^2$ characterized by $A$ and $B$ in (\ref{46}). Especially, the functions $F(k)$ and $F_0(k)$ in (\ref{68}) and (\ref{91}), respectively, will coincide with the corresponding functions for $-\Delta$, see e.g.\ \cite{BE:2008} and \cite{KostrykinSchrader:2006b}.

We remark that the transformation of the matrices $A\rightarrow A''$ and $B\rightarrow B''$ and vice versa (actually we consider below the converse direction) cannot be achieved by perturbing the negative Laplacian by a magnetic flux which corresponds to an operator acting on the edges as
\begin{equation}
\label{3000}
\left(\frac{\ud\phantom{x_j}}{\ud x_j}-\ui A_j(x_j)\right)^2\psi(x_j) \ \quad \mbox{for} \quad 1\leq j\leq E.
\end{equation}
Kostrykin and Schrader have shown in \cite{KS:2003} that this operator is related to the negative Laplacian $-\Delta$ by a unitary transformation of the corresponding S-matrices. This means that the Laplacian perturbed by a magnetic flux can also be characterized by two matrices $A$ and $B$ obeying (\ref{47}). But with a local gauge transformation this system can be transformed to a quantum graph system with the pure Laplacian which is now characterized by two new matrices $\widetilde{A}$ and $\widetilde{B}$. These new matrices are obtained by the old ones by
\begin{equation}
\label{4000}
\widetilde{A}=AU \quad \mbox{and} \quad \widetilde{B}=BU
\end{equation}
where $U$ is a diagonal unitary matrix. If we calculate the S-matrices for these systems we obtain
\begin{equation}
\label{4001}
S(\widetilde{A},\widetilde{B};k)=US(A,B;k)U^+.
\end{equation}
In particular this means that the S-matrix is $k$-independent iff the original S-matrix is $k$-independent. By a result of \cite{Schrader:2007} we conclude that in the sense of (\ref{61}) (see \cite{Kuchment:2004}) the corresponding matrix $P_{\ker B}^{\bot}LP_{\ker B}^{\bot}$ is zero iff $P_{\ker \widetilde{B}}^{\bot}\widetilde{L}P_{\ker \widetilde{B}}^{\bot}$
is zero. This feature is obviously not given by the transformation $A,B$ to $A'',B''$ especially in (\ref{72}) taking into account that the transformation $A,B$ to $A',B'$ in (\ref{57}) and the corresponding transformation $L$ to $L'$ possess this feature.

Kostrykin and Schrader have shown in \cite{KostrykinSchrader:2006} that the negative Laplacian $-\Delta$ possesses time-reversal symmetry iff $S^T=S$. Obviously, the transformation (\ref{4001}) doesn't maintain this symmetry in general.

In both cases ($H_{BK}$ and $H_{BK}^2$)  we get the same length $\mathfrak{l}_i$ for the edge $e_i$ of the quantum graph for the corresponding momentum operator or kinetic energy operator. Thus, we choose $\mathfrak{l}_i$ for the lengths of the graph and endow it with a metric structure as in section \ref{7}. 
\section{Connection between $H_{BK}$ and $H_{BK}^2$}
\label{ab8}
We also want to reveal the link between $H_{BK}$ and $H_{BK}^2$, the last one considered as the ``squared'' Berry-Keating operator as in (\ref{36}). Therefore, we are starting from $(H_{BK};A,B)$ with corresponding S-matrix $\mathcal{S}(A,B)$ and then calculate $(H_{BK}^2;\widetilde{A},\widetilde{B})$ with corresponding S-matrix $S(\widetilde{A},\widetilde{B})$. Using the definitions (\ref{20}) we obtain from (\ref{36}) the two (necessary and sufficient) equations, setting $\phi:=H_{BK}\psi$, 
\begin{equation}
\label{ab6}
\eqalign{
A\left(UD_{(\ba\bb)}^{\frac{1}{2}}\Psi_{\mathrm{bv}}\right)_1+B\left(UD_{(\ba\bb)}^{\frac{1}{2}}\Psi_{\mathrm{bv}}\right)_2 &=0\\
A\left(UD_{(\ba\bb)}^{\frac{1}{2}}\Phi_{\mathrm{bv}}\right)_1+B\left(UD_{(\ba\bb)}^{\frac{1}{2}}\Phi_{\mathrm{bv}}\right)_2 &=0.
}
\end{equation}
Making the definitions
\begin{equation}
\label{ab10}
\psi(\ba)=\bigl(\psi_1(a_1),\ldots,\psi_E(a_E)\bigl)^T \quad \mbox{and} \quad \psi(\bb)=\bigl(\psi_{1}(b_1),\ldots,\psi_{E}(b_{E})\bigr)^T
\end{equation}
the equations (\ref{ab6})can be transformed into (see definition (\ref{60}))
\begin{equation}
\label{ab11}
\eqalign{
\psi(\ba) &=\left(\frac{\bb}{\ba}\right)^{\frac{1}{2}}\mathcal{S}(A,B)\psi(\bb) \\ 
\psi'(\ba) &=\left(\frac{\bb}{\ba}\right)^{\frac{3}{2}}\mathcal{S}(A,B)\psi'(\bb).
}
\end{equation}
This is equivalent to the equation
\begin{equation}
\label{ab12}
\eqalign{
& 
\left(
\begin{array}{cc}
-\eins & \left(\frac{\ba}{\bb}\right)^{\frac{1}{2}}\mathcal{S}(A,B)\\
0 & 0
\end{array}
\right)
\D\Psi_{\mathrm{bv}}\\
& \hspace{3cm} +B
\left(
\begin{array}{cc}
0 & 0 \\
\eins & \left(\frac{\bb}{\ba}\right)^{\frac{1}{2}}\mathcal{S}(A,B)\\
\end{array}
\right)
\D\Psi_{\mathrm{bv}}'=0.
}
\end{equation}
Comparing (\ref{ab12}) with (\ref{46}), we infer for the matrices (one possible choice) $\widetilde{A}$ and $\widetilde{B}$
\begin{equation}
\label{ab13}
\widetilde{A}=
\left(
\begin{array}{cc}
-\eins & \left(\frac{\ba}{\bb}\right)^{\frac{1}{2}}\mathcal{S}(A,B)\\
0 & 0
\end{array}
\right)
,\quad 
\widetilde{B}=
\left(
\begin{array}{cc}
0 & 0 \\
\eins & \left(\frac{\bb}{\ba}\right)^{\frac{1}{2}}\mathcal{S}(A,B)
\end{array}
\right).
\end{equation}
It is a simple calculation that indeed the matrices $\widetilde{A}$ and $\widetilde{B}$ fulfil (\ref{47}). In order to calculate $S(\widetilde{A},\widetilde{B})$ we make two observations. First, we infer from (\ref{58}) and (\ref{ab13}) that $L'=0$ in the decomposition (\ref{61}). Furthermore, we notice that 
\begin{equation}
\label{ab15}
\eqalign{
{\ker \widetilde{B}'}^{\bot} &=\spa\left\{
\left(
\begin{array}{c}
\mathcal{S}(A,B)e_i\\
e_i 
\end{array}
\right)
;
\quad i=1,\ldots, E\right\}
,\\
\ker \widetilde{B}'&=\spa\left\{
\left(
\begin{array}{c}
\mathcal{S}(A,B)e_i\\
-e_i 
\end{array}
\right)
;
\quad i=1,\ldots, E\right\},
}
\end{equation}
where $\spa$ denotes the linear span of the corresponding vectors, and the vectors $e_i$ span the space $\kz^E$. Therefore, by (\ref{72}) we conclude 
\begin{equation}
\label{ab20}
L''=L'=0.
\end{equation}
Thus, by (\ref{67}) and after an easy (but lengthy) calculation we obtain for the S-matrix of $(H_{BK}^2;\widetilde{A},\widetilde{B})$
\begin{equation}
\label{ab14}
S(\widetilde{A},\widetilde{B})=
\left(
\begin{array}{cc}
0 & \mathcal{S}(A,B) \\
\mathcal{S}(A,B)^+ & 0
\end{array}
\right)
.
\end{equation}
Therefore, we have proved the following proposition.
\begin{prop}
\label{ab21}
$(H_{BK}^2;\widetilde{A},\widetilde{B})$ is the ``squared'' operator of some $(H_{BK};A,B)$ in the sense of (\ref{36}) iff the corresponding S-matrices fulfil (\ref{ab14}).
\end{prop}
We remark that a similar relation holds for the usual Laplace operator $-\Delta$ and the momentum operator $p$ on compact graphs.
\section{Trace formulae and Weyl's law}
\label{100}
We are now in the position to give explicit expressions for the behaviour of the eigenvalue counting functions for large eigenvalues and give trace formulae for the Berry-Keating operator and the ``squared'' Berry-Keating operator on compact graphs. These results are immediate consequences of sections \ref{9}, \ref{27} and the results in \cite{BE:2008}. The proofs of the claims for the Berry-Keating operator are quite analogous to \cite{BE:2008} and, therefore, we only give a short outline of some steps of the proof. Since the trace formulae differ in some details we formulate these formulae in one theorem and one corollary. First of all we introduce an appropriate space of test functions as in \cite{BE:2008}.
\begin{defn}
\label{102}
For each $r\geq 0$ the space $H_r$ consists of all functions
$h:\kz\to\kz$ satisfying the following conditions:
\begin{itemize}
\item $h$ is even, i.e., $h(k)=h(-k)$. 
\item For each $h\in H_r$ there exists $\delta>0$ such that $h$ is analytic 
in the strip $M_{r+\delta}\coloneqq \{k\in\kz; \ |\im k|<r+\delta\}$.
\item For each $h\in H_r$ there exists $\eta>0$ such that 
$h(k)=O\left(\frac{1}{(1+|k|)^{1+\eta}}\right)$ on $M_{r+\delta}$, $k\rightarrow\infty$.
\end{itemize}
\end{defn}
We denote by $k_n$ the ``energies'' respectively ``wave numbers'' of $H_{\mathrm{BK}}$  respectively $H_{\mathrm{BK}}^2$ and by $g_n$ the corresponding multiplicities which are identical with the order of the corresponding zeros $k_n$ of $\mathcal{F}$ in (\ref{75}) for $n\in \nz_0$ respectively zeros $k_n\neq0$ of $F$ in (\ref{68}) for $n\in\nz$. $n=0$ corresponds to the ``energy'' respectively ``wave number'' zero, and the energies respectively the nonnegative wave numbers are ordered with respect to their absolute value $|k_n|$ in increasing order. However, the (finitely many) imaginary wave numbers are omitted. Furthermore, we denote the self-adjoint realizations characterized by (\ref{20}) respectively (\ref{46}) by $(H_{\mathrm{BK}};A,B)$ respectively $(H_{BK}^2;\widetilde{A},\widetilde{B})$. Notice that in the first case $A,B\in\Mat(E\times E,\kz)$ whereas in the second case $\widetilde{A},\widetilde{B}\in\Mat(2E\times 2E,\kz)$. In addition we denote by $\lf_{\min}$ the minimal length of the graph with respect to the definition of section \ref{7}. The minimal positive eigenvalue of $L''$ in (\ref{72}) is denoted by ${\lambda''}_{\min}^+$ and the unique minimum of the function
\begin{equation}
\label{4103}
\lf(\kappa)\coloneqq \frac{1}{\kappa}\ln(2E)+\frac{2}{\kappa}\arth\left(\frac{\kappa}{{\lambda''}_{\min}^+}\right)
\end{equation}
by $\sigma$. For convenience, we denote the total length of the graph 
by
\begin{equation}
\label{4120}
\mathfrak{L}\coloneqq \sum\limits_{i=1}^E\lf_i.
\end{equation}
Furthermore, by a hat $\hat{\cdot}$ we denote the Fourier transform (see (\ref{308})) and $\cdot\ast\cdot$ denotes the convolution of two functions in the distributional sense, see e.g.\ \cite{Reed:1975}. For convenience, we assume that the graph $\Gamma$ is local with respect to the S-matrix $S''(A,B;k)$ which means that the scattering between two endpoints is only allowed for adjacent edge ends, see \cite{KostrykinSchrader:2006} for a precise definition. This has the effect that in the trace formula the periodic orbits are with respect to the classical topology as explained in section \ref{7}. Otherwise we must interpret the periodic orbits with respect to the topology induced by the S-matrices which will differ from the one in section \ref{7} and must then be interpreted as a quantum mechanical topology. However, in \cite{KostrykinSchrader:2006} it was shown that there exists always at least one graph, for which the S-matrix is local. In order to interpret the right side of 
(\ref{19905})
for $H_{\mathrm{BK}}$ as a sum of periodic orbits, we replace the edges by directed edges as mentioned in section \ref{7}. Furthermore, we assume that the S-matrix $\mathcal{S}(A,B)$ is local with respect to the directed edges. This means that $\mathcal{S}(A,B)_{ij}=0$ if $e_i$ and $e_j$ share no vertex $v_{ij}$ for which $e_j$ has the direction towards $v_{ij}$ and $e_i$ has the direction away from $v_{ij}$. Again, it is always possible to find such a graph. We get the following theorem for $H_{\mathrm{BK}}$.
\begin{theorem}[Trace formula for $H_{\mathrm{BK}}$]
\label{11104}
Let $\Gamma$  be a compact metric graph and $(H_{\mathrm{BK}};A,B)$ with the above assumptions be given. Let $h\in H_r$ with any $r\geq0$. Then the following trace formula holds (where $\hat{h}$ denotes the Fourier transform of $h$ defined as in eq. (\ref{308}) and $g_n$ the multiplicity of the eigenvalue $k_n$) 
\begin{equation}
\label{19905}
 \sum_{n=0}^{\infty} g _n \, h(k_n) = \phantom{+}\mathfrak{L} \, \hat{h}(0) +2\sum\limits_{\gamma\in\mathfrak{P}}\re\left(\mathcal{A}_{\gamma}\right)\hat{h}(\lf_{\gamma}).
\end{equation}
\end{theorem}
The amplitude functions $\mathcal{A}_{\gamma}$ are constructed from the S-matrix elements with respect to the periodic orbits $\gamma$, see \cite{BE:2008,KottosSmilansky:1998} for a precise definition of this construction. The proof of this theorem is quite analogous to \cite{BE:2008}. Since we have no $k$-dependence of $\mathcal{S}(A,B)$ in (\ref{25}), we can omit the requirement of the minimal length in contrast to the following corollary \ref{104}. This also leads to the simple product of the real part of the amplitude functions $\mathcal{A}_{\gamma}$ and the Fourier transform of $h$ in the identity (\ref{19905}). Furthermore, since the secular equation (\ref{23}) respectively (\ref{75}) holds also for the eigenvalue zero of $H_{\mathrm{BK}}$, the term $g_0 -\frac{1}{2}N$ does not appear in (\ref{19905}) in contrast to (\ref{105}) for $H_{\mathrm{BK}}^2$, where $N$ denotes the multiplicity of the (possible) zero $k_0=0$ of $F(k)$. For $H_{\mathrm{BK}}^2$ we get the following trace formula.
\begin{theorem}[Trace formula for $H_{\mathrm{BK}}^2$]
\label{104}
Let $\Gamma$  be a compact metric graph and $(H_{BK}^2;\widetilde{A},\widetilde{B})$ with the above assumptions be given. Let the condition $\lf_{\min}>\lf(\sigma)$ be fulfilled and let $h\in H_r$ with $r\geq\sigma$. Then the following trace formula holds
\begin{equation}
\label{105}
\fl\eqalign{ \sum_{n=0}^{\infty} g _n \, h(k_n) 
 &= \phantom{+}\mathfrak{L} \, \hat{h}(0) + \bigl(g_0 -\frac{1}{2}N \bigr)h(0)
    -\frac{1}{4\pi}\int_{-\infty}^{+\infty}h(k)\,\frac{\im\mtr S''(A,B;k)}{k}
    \ \ud k \\
 &  \quad +\sum\limits_{\gamma\in\mathfrak{P}}\left[\bigl(\hat{h}\ast\hat{A}_{\gamma}
    \bigr) (\lf_{\gamma}) + \bigl(\hat{h}\ast\hat{\overline{A}}_{\gamma}\bigr)(\lf_{\gamma}) \right]
.
}
\end{equation}
\end{theorem}
Again, the amplitude functions $A_{\gamma}$ are constructed from the S-matrix elements with respect to the periodic orbits $\gamma$ and $g_0$ denotes the multiplicity of the eigenvalue one of $S''(A,B;k')C(\ba,\bb;k')$ for any $k'\in\rz\setminus\{0\}$ (see section \ref{90}).

Since we have previously seen that the spectrum of $(H_{BK}^2;\widetilde{A},\widetilde{B})$ coincides with some self-adjoint realization of $-\Delta$ on the graph by adapting the lengths and with the results in \cite{BE:2008}, we get Weyl's law:
\begin{theorem}[Weyl's law for $H_{\mathrm{BK}}^2$]
\label{110}
Given the eigenvalues of some $(H_{BK}^2;\widetilde{A},\widetilde{B})$ in increasing order denoted by $\lambda_n=k_n^2$. Then for the counting function $N(\lambda)\coloneqq \#\left\{n; \ k_n^2\leq\lambda\right\}$ the following asymptotic law holds 
\begin{equation}
\label{4500}
N(\lambda) \sim\frac{\mathfrak{L}}{\pi}\sqrt{\lambda} \quad \mbox{for} \quad\lambda\rightarrow\infty.
\end{equation}
\end{theorem}
The same asymptotic law holds for $(H_{BK};A,B)$ replacing $\lambda$ by $k$ on the left-hand side and replacing $\sqrt{\lambda}$ by $k$ at the right-hand side on the equation, i.e. we have
\begin{theorem}[Weyl's law for $H_{\mathrm{BK}}$]
\label{m1}
Given the positive eigenvalues of some $\left(H_{\mathrm{BK}};A,B\right)$ in increasing order denoted by $\lambda_n=k_n$. Then for the counting function $N(k)\coloneqq \#\left\{n; \ k_n\leq k\right\}$ the following asymptotic law holds
\begin{equation}
\label{m2}
N(k)\sim\frac{\mathfrak{L}}{\pi}k \quad \mbox{for} \quad k\rightarrow\infty.
\end{equation}
\end{theorem}
The theorem \ref{m1} can be proved by a suitable Karamata-Tauberian theorem (see e.g. \cite{Karamata:1931} and \cite{BE:2008}) or alternatively by applying theorem \ref{110} to the ``squared'' operator $H_{BK}^2$ of $H_{BK}$ in the sense of \eref{36} (the spectrum of the eigenvalues of $H_{BK}$ and the wave numbers of $H_{BK}^2$ coincides then and therefore the corresponding eigenvalue respectively wave number counting functions are the same). Comparing the theorems \ref{110} and \ref{m1} with the asymptotics of the counting function for the nontrivial Riemann zeros \eref{2}, we therefore can conclude:
\begin{theorem}[No-go theorem]
\label{150}
Neither $H_{\mathrm{BK}}$ nor $H_{\mathrm{BK}}^2$ yields as eigenvalues the nontrivial Riemann zeros if these are self-adjoint realizations on any compact graph.
\end{theorem}
\section{Simple examples}
\label{120}
We shall give a simple example for a wave packet and its time-evolution with respect to the Berry-Keating operator in $\mathcal{H}=L^2(\rz_>,\ud x)$ discussed in section \ref{26}. Furthermore, we give an example for a realization of $H_{\mathrm{BK}}$ and $H_{\mathrm{BK}}^2$ on the simplest construction of a graph which consists of a single edge. Finally, we present some trace formulae for the presented examples.
%
%
\begin{ex}
\label{e2}
For $\psi(x,0)=\phi(x)$ in (\ref{313}) we define ($x\in\rz_{>}$)
\begin{equation}
\label{701}
\phi(x)\coloneqq \frac{\alpha}{\ue^x+1} \quad \mbox{with} \quad \alpha=\frac{1}{\sqrt{\ln 2-\frac{1}{2}}}.
\end{equation}
(With this choice for $\alpha$ it holds $\|\phi\|=1$.) From (\ref{310}) we obtain
\begin{equation}
\label{6000}
\psi(x,t)=(U(t)\phi)(x)=\frac{\alpha\ue^{-\frac{t}{2}}}{\ue^{x\ue^{-t}}+1} \quad \mbox{with} \quad t\in\rz.
\end{equation} 
Thus, we get the large-$t$ asymptotics
\begin{equation}
\label{6001}
\psi\sim\frac{\alpha}{2}\ue^{-\frac{t}{2}} \quad \mbox{for} \quad t\rightarrow\infty.
\end{equation}
On the other hand, with
\begin{equation}
\label{6002}
K_\mathrm{BK}(x,x_0;t)= \int\limits_{-\infty}^{\infty}\psi_k(x)\overline{\psi}_k(x_0)\ue^{-\ui k t}\ud k,
\end{equation}
(\ref{480}) and (\ref{313}), we get
\begin{equation}
\label{485}
\psi(x,t)=\int\limits_{-\infty}^{\infty}A(k)\psi_k(x)\ue^{-\ui k t} \ud k.
\end{equation}
A direct calculation using (\ref{2001}) and the integral representation of $\zeta(s)$ as a Mellin transform (see $\cite[p.\,20]{Magnus:1966}$) yields 
\begin{equation}
\label{490}
A(k)=\frac{\alpha}{\sqrt{2\pi}}\left(1-\sqrt{2}\, 2^{\ui k}\right)\Gamma\left(\frac{1}{2}-\ui k\right)\zeta\left(\frac{1}{2}-\ui k\right).
\end{equation}
With (see $\cite[p.\,13]{Magnus:1966}$)
\begin{equation}
\label{6005}
\left|\Gamma\left(\frac{1}{2}-\ui k\right)\right|\sim\sqrt{2\pi}\ue^{-\frac{\pi}{2}|k|} \quad \mbox{for} \quad k\in\rz, \quad |k|\rightarrow\infty,
\end{equation}
we get for the large-$k$ asymptotics of $|A(k)|^2$
\begin{equation}
\label{6006}
\eqalign{
|A(k)|^2 & \sim\alpha^2\left(3-2\sqrt{2}\cos(k\ln 2)\right)\ue^{-\pi|k|}\left|\zeta\left(\frac{1}{2}-\ui k\right)\right|^2 \\
& \hspace{5mm} \mbox{for} \quad k\in\rz, \quad |k|\rightarrow\infty,
}
\end{equation}
which gives a sufficient condition for $A\in L^2(\rz,\ud k)$. If we consider the continuous representation (\ref{485}) of $\psi(x,t)$, we see that $\psi(x,t)$ gets  no contribution from the wave packet $A(k)$ exactly at the wave numbers $k$ corresponding to the conjectured nontrivial Riemann zeros. This is reminiscent to the absorption spectrum interpretation of the nontrivial Riemann zeros by Connes $\cite{Connes:1996,Connes:1999}$, but of course reveals no insight to the position of the nontrivial Riemann zeros.
\end{ex}

\begin{ex}
\label{50000}
For a single edge $I=[a,b]$ (one-dimensional quantum billiard) the matrices $A$ and $B$ are arbitrary numbers fulfilling (\ref{18}). The equations (\ref{20}) and (\ref{25}) lead then with 
\begin{equation}
\label{30002}
\mathcal{S}(A,B)=:\ue^{-2\pi\ui c}
\end{equation}
to 
\begin{equation}
\label{7300}
\eqalign{
\psi(a)&= \mathcal{S}(A,B)\sqrt{\frac{b}{a}}\psi(b) \\
       &= \sqrt{\frac{b}{a}}\ue^{-2\pi\ui c}\psi(b) \quad \mbox{with} \quad c\in[0,1).
}
\end{equation}
The eigenvalue spectrum is given by
\begin{equation}
\label{140}
k_n=\frac{2\pi}{\ln \frac{b}{a}}(n+c) \quad \mbox{with} \quad c\in[0,1) \quad \mbox{and} \quad n\in \gz.
\end{equation}
We now want to calculate $H_{\mathrm{BK}}^2$ as defined in (\ref{36}) with (\ref{7300}), in particular the S-matrix and then compare it with the results in section \ref{ab8}. In order to distinguish the characterizing matrices, we denote these with the subscript $\boldsymbol{\cdot}_{H_{\mathrm{BK}}}$ and $\boldsymbol{\cdot}_{H_{\mathrm{BK}}^2}$. First, we derive the transformation of $A_{H_{\mathrm{BK}}},B_{H_{\mathrm{BK}}}$ into $A_{H_{\mathrm{BK}}^2},B_{H_{\mathrm{BK}}^2}$ for the corresponding operators related by (\ref{36}). We get the additional condition
\begin{equation}
\label{6007}
\psi'(a)=\left(\frac{b}{a}\right)^{\frac{3}{2}}\ue^{-2\pi\ui c}\psi'(b) \quad \mbox{with} \quad c\in[0,1).
\end{equation}
The two conditions are equivalent to
\begin{equation}
\label {130}
\fl\eqalign{
0 &= 
\left(
\begin{array}{cc}
-1 & \left(\frac{b}{a}\right)^{\frac{1}{2}}\ue^{-2\pi\ui c} \\
0 & 0
\end{array}
\right)
\Psi_{\mathrm{bv}}+
\left(
\begin{array}{cc}
0 & 0 \\
1 & \left(\frac{b}{a}\right)^{\frac{3}{2}}\ue^{-2\pi\ui c}
\end{array}
\right)
\Psi_{\mathrm{bv}}'
\\
\Leftrightarrow \ 0 &= 
\left(
\begin{array}{cc}
-1 & \left(\frac{b}{a}\right)^{\frac{1}{2}}\ue^{-2\pi\ui c} \\
0 & 0
\end{array}
\right)
\left(
\begin{array}{cc}
\frac{1}{a} & 0 \\
0 & \frac{1}{b}
\end{array}
\right)
D_{(ab)}
\Psi_{\mathrm{bv}}\\
&\hspace{3cm} 
+\left(
\begin{array}{cc}
0 & 0 \\
1 & \left(\frac{b}{a}\right)^{\frac{3}{2}}\ue^{-2\pi\ui c}
\end{array}
\right)
\left(
\begin{array}{cc}
\frac{1}{a} & 0 \\
0 & \frac{1}{b}
\end{array}
\right)
D_{(ab)}
\Psi_{\mathrm{bv}}'
\\
\Leftrightarrow  \ 0 &= 
\left(
\begin{array}{cc}
-1 & \left(\frac{a}{b}\right)^{\frac{1}{2}}\ue^{-2\pi\ui c} \\
0 & 0
\end{array}
\right)
D_{(ab)}
\Psi_{\mathrm{bv}}+
\left(
\begin{array}{cc}
0 & 0 \\
1 & \left(\frac{b}{a}\right)^{\frac{1}{2}}\ue^{-2\pi\ui c}
\end{array}
\right)
D_{(ab)}
\Psi_{\mathrm{bv}}'.
}
\end{equation}
Therefore, we define 
\begin{equation}
\label{6008}
A_{H_{\mathrm{BK}}^2}\coloneqq 
\left(
\begin{array}{cc}
-1 & \left(\frac{a}{b}\right)^{\frac{1}{2}}\ue^{-2\pi\ui c} \\
0 & 0
\end{array}
\right)
\quad \mbox{and} \quad 
B_{H_{\mathrm{BK}}^2}\coloneqq 
\left(
\begin{array}{cc}
0 & 0 \\
1 & \left(\frac{b}{a}\right)^{\frac{1}{2}}\ue^{-2\pi\ui c}
\end{array}
\right)
\end{equation}
and recognize that indeed
\begin{equation}
\label{6500}
A_{H_{\mathrm{BK}}^2}B_{H_{\mathrm{BK}}^2}^+=B_{H_{\mathrm{BK}}^2}A_{H_{\mathrm{BK}}^2}^+=0 \quad \mbox{and} \quad \rank\left(A_{H_{\mathrm{BK}}^2},B_{H_{\mathrm{BK}}^2}\right)=2
\end{equation}
is fulfilled. By a comparison of (\ref{130}) with (\ref{46}), we infer that $A_{H_{\mathrm{BK}}^2}$ and $B_{H_{\mathrm{BK}}^2}$ are two possible matrices to characterize $H_{\mathrm{BK}}^2$ in the sense of (\ref{46}). For $S''\bigl(A_{H_{\mathrm{BK}}^2},B_{H_{\mathrm{BK}}^2};k\bigr)$, we get
\begin{equation}
\label{714}
S''\bigl(A_{H_{\mathrm{BK}}^2},B_{H_{\mathrm{BK}}^2};k\bigr)=
\left(
\begin{array}{cc}
0 & \ue^{-2\pi\ui c}\\
\ue^{2\pi\ui c} & 0
\end{array}
\right)
=
\left(
\begin{array}{cc}
0 & \mathcal{S}(A,B)\\
\mathcal{S}(A,B)^+& 0
\end{array}
\right)
\end{equation}
in complete agreement with (\ref{ab14}) and for the secular equation (\ref{68})
\begin{equation}
\label{6502}
0=\left(\ue^{\ui\left(k\ln\left(\frac{b}{a}\right)+2\pi c\right)}-1\right)\left(\ue^{\ui\left(k\ln\left(\frac{b}{a}\right)-2\pi c\right)}-1\right).
\end{equation}
This leads to the ``wave numbers''
\begin{equation}
\label{700}
k_n=\frac{2\pi}{\ln \frac{b}{a}}(n\pm c), \quad n\in\gz
\end{equation}
with $c$ as in (\ref{140}). Obviously, with (\ref{700}) and (\ref{140}) Weyl's law is fulfilled even for small $n$.
Alternatively, since these are from the classical point of view integrable systems we can perform an EBK-quantization for $H_{\mathrm{BK}}$ and $H_{\mathrm{BK}}^2$
$\cite{Einstein:1917,Keller:1958}$. In this semiclassical quantization rule the spectrum consists of energies $E_n$ (for convenience we use  now the same letter $E_n$ for $k_n$ respectively $\lambda_n$ as in the sections \ref{26} respectively \ref{7680}) for which ($\hbar=1$)
\begin{equation}
\label{710}
I_n(E_{n})=\left(n+\frac{\mu_n}{4}\right) \quad \mbox{with} \quad n\geq 0  
\end{equation}
is fulfilled. Therein $\mu_n$ denotes the so-called Maslov index and
\begin{equation}
\label{711}
I_n(E_{n})=\frac{1}{2\pi}\int\limits_{\gamma_n}p\ud x
\end{equation}
is the classical action of a periodic orbit $\gamma_n$ which is a subset of the hypersurface $H_{\mathrm{cl}}=E_{n}$ respectively $\widetilde{H}_{\mathrm{cl}}=E_{n}$. For $H_{\mathrm{cl}}$ in (\ref{201}) with the ring system structure mentioned in section \ref{599}, we get from (\ref{710})
\begin{equation}
\label{712}
E_{n}=\frac{2\pi}{\ln\left(\frac{b}{a}\right)}\left(n+\frac{\mu_n}{4}\right), 
\end{equation}
and for $\widetilde{H}_{\mathrm{cl}}$ in (\ref{31}) (also with the ring system structure)
\begin{equation}
\label{713}
\sqrt{E_{n}}=k_n=\frac{2\pi}{\ln\left(\frac{b}{a}\right)}\left(n+\frac{\mu_n}{4}\right).
\end{equation}
\end{ex}
A comparison of (\ref{712}) with (\ref{140}) yields for the Maslov indices $\mu_n=4c$ for $H_{\mathrm{cl}}$. For $\widetilde{H}_{\mathrm{BK}}$ we get two Maslov indices, $\mu_n=4c$ for $n=0,2,4,\ldots$ and $\mu_n=-4c$ for $n=1,3,5,\ldots$. Since a Maslov index is at most defined modulo $4$ and because of $c\in [0,1)$, the above second Maslov indices $\mu_n=-4c$ correspond to the Maslov indices $\tilde{\mu}_n=4(1-c)$ for $n=1,3,\ldots$. We stress that the EBK-quantization for $\widetilde{H}_{\mathrm{cl}}$ with a classical ``hard wall'' boundary condition yields
\begin{equation}
\label{6600}
\sqrt{E_{n}}=k_n=\frac{\pi}{\ln\left(\frac{b}{a}\right)}\left(n+\frac{\mu_n}{4}\right),
\end{equation}
which differs from (\ref{713}) by a factor $2$. We mention that the S-matrix elements of $S''\bigl(A_{H_{\mathrm{BK}}^2},B_{H_{\mathrm{BK}}^2};k\bigr)$ for Dirichlet ($D$), Neumann ($N$) or Robin ($R$) boundary conditions are given by
\begin{equation}
\label{6580}
\eqalign{
S''\left(A_{H_{\mathrm{BK}}^2},B_{H_{\mathrm{BK}}^2};k\right)_{ij} &=
\cases{
-\delta_{ij} & for $D$,\\
\delta_{ij} & for $N$, \\
-\delta_{ij}\frac{\rho_j-\ui k}{\rho_j +\ui k} & for $R$,\\
}
\\
& \mbox{with} \ \rho_j\in\rz \ \mbox{and} \ i,j\in\left\{1,2\right\},
}
\end{equation}
which obviously differ from (\ref{714}) and, therefore, cannot originate from a ``squared'' Berry-Keating operator. If we impose Dirichlet boundary conditions at both interval ends, we get for $F(k)$ in (\ref{68})
\begin{equation}
\label{3850}
F(k)=1-\ue^{2\ui k \ln\left(\frac{b}{a}\right)}
\end{equation}
and thus we obtain for the wave numbers $k_n$
\begin{equation}
\label{3851}
k_n=\frac{\pi}{\ln\left(\frac{b}{a}\right)}n, \quad n\in \gz\setminus\{0\},
\end{equation}
wherein we have taken into account that $\lambda=k_0^2=0$ is not an eigenvalue for the Dirichlet case. In contrast to the Dirichlet case, $\lambda=k_0=0$ is an eigenvalue for Neumann boundary conditions at both interval ends, and the nonzero wave numbers coincide with the Dirichlet case (\ref{3851}). For Robin boundary conditions at both interval ends, we get
\begin{equation}
\label{3852}
F(k)=1-\frac{\left(\rho_1-\ui k\right)\left(\rho_2-\ui k\right)}{\left(\rho_1+\ui k\right)\left(\rho_2+\ui k\right)}\ue^{2\ui k \ln\left(\frac{b}{a}\right)}.
\end{equation}
Again, by a comparison of (\ref{3851}) with (\ref{6600}), we conclude for the Maslov indices for pure Dirichlet and Neumann boundary conditions 
\begin{equation}
\label{3853}
\mu_n=0 \quad \mbox{for} \quad n\in \nz
\end{equation}
and additionally $\mu_0=0$ for the Neumann case. For the Robin boundary conditions on both interval ends the Maslov indices have to be individually calculated for each $n\in \nz_0$ by (\ref{3852}).

The above considerations underline the fact that the form of the S-matrix (\ref{714}) corresponds to a pure ring system as in the case of the negative Laplacian $-\Delta$, see $\cite{KostrykinSchrader:2006}$. The occurrence of possible noninteger ``Maslov indices'' originates simply from the fact that we have a discontinuous crossover by turning once around in the ring system (one-dimensional torus) in contrast to the ``usual'' continuity requirement of the wave function, see e.g.\ $\cite{Stöckmann:1999}$.
\begin{ex}
We shall present a trace formula for the time-evolution operator $U(t)$ in (\ref{309}) and (\ref{310}) for $H_{\mathrm{BK}}$ acting on a single edge with the assigned interval $I=[1,b]$:
\begin{equation}
\label{20000}
\left(U(t)\phi\right)(x) \coloneqq  \sum\limits_{n=-\infty}^{\infty} \psi_n(x)\left<\psi_n,\phi\right>\ue^{-\ui k_nt}, \quad \phi\in L^2(I,\ud x),
\end{equation}
with the eigenvalues $k_n=\frac{2\pi}{\ln b} (n+c)$, $n\in\gz$, $c\in[0,1)$, and the normalized eigenfunctions
\begin{equation}
\label{20001}
\psi_n(x)=\frac{1}{\sqrt{x\ln b}}\ue^{\ui k_n \ln x} \quad \mbox{with} \quad n\in\gz.
\end{equation}
For the corresponding (not retarded) integral kernel of $U(t)$ we get by $\cite[p.\,20]{Gelfand:1960}$ (in a distributional sense acting on $\mathcal{D}(I)\subset L^2(I,\ud x)$ identified by the continuous representatives; $g(x,x_0;t)\coloneqq \frac{2\pi}{\ln b}\left[\ln x-\ln x_0 -t\right]$)
\begin{equation}
\label{20003}
\eqalign{
K(x,x_0;t) &\coloneqq  \sum\limits_{n=-\infty}^{\infty} \psi_n(x)\overline{\psi}_n(x_0)\ue^{-\ui k_nt} = \frac{\ue^{\ui c g(x,x_0;t)}}{\sqrt{xx_0}\ln b}\sum\limits_{n=-\infty}^{\infty}\ue^{\ui g(x,x_0;t) n}\\
           &= \frac{2\pi}{\ln b}\frac{\ue^{\ui c g(x,x_0;t)}}{\sqrt{xx_0}}\sum\limits_{n=-\infty}^{\infty}\delta(g(x,x_0;t)+2\pi n) \\
           &= \ue^{\ui c g(x,x_0;t)}\sum\limits_{n=-\infty}^{\infty}b^{\frac{n}{2}}\ue^{\frac{t}{2}}\delta\left(xb^n-x_0\ue^t\right).
}
\end{equation}
If we take the trace of $U(t)$, we obtain with (\ref{20003}) (by defining the ``period'' $T=\ln b$ [see (\ref{4})] and the Maslov index $\mu\coloneqq 4c$)
\begin{equation}
\label{20005}
\eqalign{
\trr U(t) &\coloneqq \int\limits_{1}^{b}K(x,x;t)\ud x = \sum\limits_{n=-\infty}^{\infty}\ue^{-\ui k_n t} \\
             &= T\sum\limits_{n=-\infty}^{\infty}\ue^{-\ui \frac{\pi}{2} \mu n}\delta\left(t-nT\right).
}
\end{equation}
If we now choose a test function $h$ of $H_r$ (definition \ref{102}) with an arbitrary $r>0$, we get by the identity (\ref{20005}) and the symmetry of $h$ the trace formula 
\begin{equation}
\label{20006}
\eqalign{
\int\limits_{-\infty}^{\infty}\hat{h}(t)\trr U(t)\ud t &= \sum\limits_{n=-\infty}^{\infty}h(k_n)\\
                                                          &= T\hat{h}(0) + T \sum\limits_{n=1}^{\infty}\left(\ue^{-\ui \frac{\pi}{2} \mu n}\hat{h}(nT)+\ue^{\ui \frac{\pi}{2} \mu n}\hat{h}(-nT)\right)\\
                                                          &= T\hat{h}(0) + 2T \sum\limits_{n=1}^{\infty}\cos\left(\frac{\pi}{2} \mu n\right)\hat{h}(nT).
}
\end{equation}
We recall that the S-matrix for this quantum graph is $\mathcal{S}(A,B)=\ue^{-2\pi\ui c}=\ue^{-\ui \frac{\pi}{2} \mu}$ (see (\ref{30002})), and the length of the (single) edge is $\mathfrak{l}=\mathfrak{L}=\ln \frac{b}{1}=\ln b$. Since we have a directed edge, there is only one possibility for the orientation of the periodic orbits and, therefore, the periodic orbits can be labelled by the natural numbers, and the corresponding lengths of the periodic orbits are $\mathfrak{l}_n=n\ln b$ and all are multiples of one primitive periodic orbit with length $\mathfrak{l}_1=\ln b=T$. For the amplitude functions we get (see $\cite{BE:2008}$) $\mathcal{A}_n=\mathfrak{l}_1\ue^{-2\pi\ui cn}=T\ue^{-\ui \frac{\pi}{2} \mu n}$. Applying (\ref{19905}) we get (\ref{20006}), which confirms the trace formula in theorem \ref{11104}.
\end{ex}
\begin{ex}
We shall present a trace formula for the kernel $\wK(x,x_0;t)$ of the unitary evolution operator $\ue^{-\ui tH_{\mathrm{BK}}^2}$ of $H_{\mathrm{BK}}^2$ with Dirichlet boundary conditions $(D)$ on a single edge $e$ with assigned interval $I=[1,b]$. The eigenvalues are given by (\ref{3851}), thus the (Feynman-)kernel reads
\begin{equation}
\label{b1}
\wK(x,x_0;t)\coloneqq {\sum\limits_{n=1}^{\infty}}\psi_n(x)\overline{\psi}_n(x_0)\ue^{-\ui k_n^2t}
\end{equation}
with the normalized eigenfunctions 
\begin{equation}
\label{b2}
\psi_n(x)\coloneqq \sqrt{\frac{2}{\mathfrak{l}}}\,\frac{\sin\left(n \pi\mdfrac{\ln x}{\mathfrak{l}}\right)}{\sqrt{x}}, \quad n\in\nz,
\end{equation}
where $\mathfrak{l}\coloneqq \ln b$ denotes the length of the edge $e$. Using a suitable addition theorem for trigonometric functions, we get two alternative expressions for $\wK(x,x_0;t)$ (see $\cite[p.\,371]{Magnus:1966})$
\begin{equation}
\label{b3}
\fl\eqalign{
 \wK(x,x_0;t) &=\frac{1}{\sqrt{xx_0}\mathfrak{l}_{\gamma_p}}\left[\Theta_3\left(\frac{1}{\mathfrak{l}_{\gamma_p}}\ln\left(\frac{x}{x_0}\right),-\frac{4\pi}{\mathfrak{l}_{\gamma_p}^2}t\right)-\Theta_3\left(\frac{1}{\mathfrak{l}_{\gamma_p}}\ln\left(xx_0\right),-\frac{4\pi}{\mathfrak{l}_{\gamma_p}^2}t\right)\right]
\\
              &=\frac{1}{2\sqrt{xx_0}\sqrt{\ui\pi t}}\sum\limits_{n=0}^{\infty}\epsilon_n\left[\ue^{\ui 2n\pi }\exp\left(\ui\frac{\left(\ln x-\ln x_0 +n\mathfrak{l}_{\gamma_p}\right)^2}{4t}\right)\right.\\
              &\hspace{3cm} \left.+\ue^{\ui(2n+1)\pi}\exp\left(\ui\frac{\left(\ln x+\ln x_0 +n\mathfrak{l}_{\gamma_p}\right)^2}{4t}\right)\right.\\
              &\hspace{3cm}+\left.\ue^{\ui 2n\pi }\exp\left(\ui\frac{\left(\ln x-\ln x_0 -n\mathfrak{l}_{\gamma_p}\right)^2}{4t}\right)\right.\\
              &\hspace{3cm}\left.+\ue^{\ui(2n+1)\pi}\exp\left(\ui\frac{\left(\ln x+\ln x_0 -n\mathfrak{l}_{\gamma_p}\right)^2}{4t}\right)\right],
}
\end{equation}
where $\Theta_3(z,\tau)$ denotes the Jacobi theta function and we have defined 
\begin{equation}
\label{c1}
\epsilon_n\coloneqq 
\cases{
\frac{1}{2} & for $n=0$ \\
1 & for $n>0$.\\ 
}
\end{equation}
$\mathfrak{l}_{\gamma_p}\coloneqq 2\mathfrak{l}=2\ln b$ is the length of the primitive periodic orbit $\gamma_p$ of the corresponding classical system. Notice that the summands in the second identity in (\ref{b3}) can be interpreted as contributions of free particle kernels at a fixed time $t$ corresponding to the four types of paths $p(x_0,x)$ (see section \ref{7}) joining $x_0$ and $x$ (see e.g.\ $\cite{Schrader:2007,Gaspard:2002}$). For this reason, we define (cf.\ (\ref{7703}) and $\cite[p.\,30]{Steiner:1998})$
\begin{equation}
\label{b4}
\wK_{p}(x,x_0;t)\coloneqq \frac{1}{2\sqrt{\ui\pi t}}\exp\left(\ui\frac{\mathfrak{l}_p(x_0,x)^2}{4t}\right)
\end{equation}
where $\mathfrak{l}_p(x_0,x)$ is the length of the path $p(x_0,x)$ (see (\ref{80})), and we then get by (\ref{b3}) 
\begin{equation}
\label{b5}
\eqalign{
\wK(x,x_0;t) &= \frac{1}{\sqrt{xx_0}}\sum\limits_{p(x_0,x)}\exp\left(\ui \pi n_{p(x_0,x)}\right)\wK_{p}(x,x_0;t)\\
             &= \frac{1}{\sqrt{xx_0}}\sum\limits_{p(x_0,x)}\exp\left(-\ui \frac{\pi\mu_{p(x_0,x)}}{2}\right)\wK_{p}(x,x_0;t),
}
\end{equation}
where the sum comprises all possible paths $p(x_0,x)$ joining $x_0$ and $x$, and $n_{p(x_0,x)}$ is defined as the number of reflections of the path $p(x_0,x)$ at the ``hard wall'' interval endpoints $1$ and $\ln b$. $\mu_{p(x_0,x)}$ denotes the Maslov index of the path $p(x_0,x)$ which is given by $\mu_{p(x_0,x)}=2n_{p(x_0,x)} \mod 4$ in agreement with the ``usual'' Maslov index for the one-dimensional billiard system corresponding to the negative Laplacian (see $\cite{Stöckmann:1999}$), and with (\ref{3853}) (in (\ref{3853}) the Maslov index corresponds to periodic orbits).
\end{ex}
\begin{ex}
Finally, we shall present an explicit trace formula (heat kernel) for a single edge with assigned interval $I=[1,b]$ for $H_{\mathrm{BK}}^2$ with Dirichlet boundary conditions $(D)$. We calculate the trace of the heat kernel of $\ue^{-tH_{\mathrm{BK}}^2}$ (replacing $t$ by $-\ui t$ in (\ref{b1}))
\begin{equation}
\label{a2}
\widetilde{K}_h(x,x_0;t):=\widetilde{K}(x,x_0;-\ui t)=\sum\limits_{k_n}\psi(x)\overline{\psi}_n(x_0)\ue^{-k_n^2t}
\end{equation}
directly and then compare the result with the trace formula (\ref{105}). Therefore, we recall that the wave numbers of $H_{\mathrm{BK}}^2$ with $(D)$ are explicitly given by (\ref{3851}). Thus, we obtain for the trace of the heat kernel (setting $\mathfrak{L}\coloneqq \frac{1}{2}\mathfrak{l}_{\gamma_p}\coloneqq \ln b$ and $\cite[p.\,371]{Magnus:1966}$)
\begin{equation}
\label{a1}
\eqalign{
\trr \ue^{-t H_\mathrm{BK}^2} & =\int\limits_1^b\sum\limits_{k_n}\psi(x)\overline{\psi}_n(x)\ue^{-k_n^2t}\ud x
                                =\sum\limits_{k_n}\ue^{-k_n^2t}\\
                              & =\frac{1}{2}\left(\Theta_3\biggl(0,\ui\frac{4\pi}{\mathfrak{l}_{\gamma_p}^2} t\biggr)-1\right)\\
                              & = \frac{\mathfrak{l}_{\gamma_p}}{4\sqrt{t\pi}}\left(\sum\limits_{n=-\infty}^{\infty}\exp\biggl(-{\frac{n^2\mathfrak{l}_{\gamma_p}^2}{4 t}}\biggr)\right)-\frac{1}{2} \\
                              & =\frac{\mathfrak{L}}{2\sqrt{\pi t}}-\frac{1}{2}+\sum\limits_{n=1}^{\infty}\frac{\mathfrak{l}_{\gamma_p}}{2\sqrt{\pi t}}\ue^{-\frac{\left(n\mathfrak{l}_{\gamma_p}\right)^2}{4t}}.
}
\end{equation}
Notice that the sums in (\ref{a1}) are absolutely convergent whereas in (\ref{20005}) the sums are convergent in the topology of $\mathcal{D}'(\rz)$ (in a distributional sense). In order to compare this result with (\ref{105}), we recall that $C(1,\ln b;k')$ and the S-matrix $S_{(D)}$ for the Dirichlet case is given by (see (\ref{91}) and (\ref{6580}))
\begin{equation}
\label{a3}
S''_{(D)}=-\eins_{2\times2} \quad \mbox{and} \quad C(1,\ln b;k')=
\left(
\begin{array}{cc}
\frac{\ln b}{2\frac{\ui}{k'} + \ln b} & \frac{2\frac{\ui}{k'}}{2\frac{\ui}{k'} + \ln b}\\
\frac{2\frac{\ui}{k'}}{2\frac{\ui}{k'} + \ln b} & \frac{\ln b}{2\frac{\ui}{k'} + \ln b}
\end{array}
\right)
.
\end{equation}
It is a simple calculation that the multiplicity $g_0$ of the eigenvalue one of $S''_{(D)}C(1,\ln b;k')$ is $g_0=0$ for any $k'\in\rz\setminus\{0\}$. Furthermore, it is obvious that the order $N$ of the zero with wave number $k_0=0$ of $F(k)$ in (\ref{68}) is $N=1$, thus $g_0 -\frac{1}{2}N =-\frac{1}{2}$. The multiplicities of the wave numbers $k_n$ are $g_n=1$ for $n\in\nz$. Since a Dirichlet boundary condition corresponds to the classical ``hard wall'' boundary condition, we conclude that the periodic orbits $\gamma$ are given by all multiples of one primitive periodic orbit $\gamma_p$ with primitive periodic orbit length $\mathfrak{l}_{\gamma_p}=2\ln b$. For the amplitude functions $A_{\gamma}$ in (\ref{105}) we obtain $A_{\gamma}=\frac{1}{2}\mathfrak{l}_{\gamma_p}$ (see $\cite{BE:2008}$). Furthermore, it is obvious that $\im S_{D}=0$. Using the test function $h(k)\coloneqq \ue^{-k^2t}$ we obtain the Fourier transform $\hat{h}(x)=\frac{1}{2\sqrt{\pi t}}\ue^{-\frac{x^2}{4t}}$. Inserting these quantities in the trace formula (\ref{105}) we get the trace formula (\ref{a1}), which again confirms the trace formula (\ref{105}). We remark that from the small-$t$ asymptotics (\ref{a1}) one obtains directly the Weyl asymptotics (\ref{4500}) using a proper Karamata-Tauberian theorem.
\end{ex}

\section{Summary and conclusions}
\label{e1}
We have studied the quantization of the extraordinarily simple classical Hamiltonian $H_{\mathrm{cl}}(x,p)=xp$ about which Berry and Keating \cite{Berry:1999b,Berry:1999} speculated that some quantization of it might yield the hypothetical Hilbert-Polya operator \cite{Weil,Weil:1999,Weil:1952,Od,Berry:1986,Connes:1996,Connes:1999} possessing  as eigenvalues the nontrivial Riemann zeros. Two quantum Hamiltonians respectively Schr\"odinger operators have been considered: The original Berry-Keating operator $H_{\mathrm{BK}}:= -\ui\hbar\left(x\frac{\ud\phantom{x}}{\ud x}+\frac{1}{2}\right)$ obtained from $H_{\mathrm{cl}}$ by Weyl ordering, and the so-called ``squared'' Berry-Keating operator $H_{\mathrm{BK}}^2:= -x^2\frac{\ud^2\phantom{x}}{\ud x^2}-2x\frac{\ud\phantom{x}}{\ud x}-\frac{1}{4}$ which is a special case of the famous Black-Scholes operator \cite{Black:1973,Merton:1973} used in the financial theory of option pricing.

In section \ref{26}, we have given a complete description of the quantum dynamics generated by $H_{\mathrm{BK}}$ acting in the Hilbert space $L^2(\rz_>, \ud x)$. While the one-dimensional quantum system governed by $H_{\mathrm{BK}}$ possesses many interesting properties, one of our main results of section \ref{26} is that the spectrum of $H_{\mathrm{BK}}$ is purely continuous corresponding to scattering states. Since there are no bound states corresponding to a discrete spectrum, it is obvious that this specific quantization of the Berry-Keating operator cannot possess the Riemann zeros as part of its spectrum. Let us point out, however, that in the simple example \ref{e2} we have studied the quantum dynamics of $H_{\mathrm{BK}}$ for the particular square-integrable wave function \eref{6000} for which it turns out that the spectral decomposition consists of a continuous wave number spectrum, $k\in\rz$, into which there are embedded infinitely many absorption lines located exactly at the wave numbers $k_n=\tau_n\in\rz$ corresponding to the nontrivial Riemann zeros  satisfying the Riemann hypothesis. To our knowledge, there is, however, no relation of this occurrence of the Riemann zeros to the absorption spectrum interpretation of Connes \cite{Connes:1996,Connes:1999}.

Analogous results have been obtained in section \ref{7680} for the ``squared'' Berry-Keating operator $H_{\mathrm{BK}}^2$ acting in the Hilbert space $L^2(\rz_>, \ud x)$. We have proved that in this case the spectrum is purely continuous too and thus there holds again a ``no-go theorem'' with respect to the identification of $H_{\mathrm{BK}}^2$ with the hypothetical Hilbert-Polya operator.

In the main part of our paper, we have dealt with the quantum dynamics of $H_{\mathrm{BK}}$ respectively $H_{\mathrm{BK}}^2$ on compact quantum graphs introduced in section \ref{7}. After having defined the Berry-Keating operator $H_{\mathrm{BK}}$ on compact graphs in section \ref{9}, we have given in proposition \ref{ab1} a complete classification of all self-adjoint extensions of $H_{\mathrm{BK}}$ on compact quantum graphs in terms of two matrices $A$ and $B$ satisfying the conditions \eref{18}. In proposition \ref{ab2} we have established a secular equation valid for any self-adjoint realization in form of a determinant whose zeros determine the discrete spectrum of $H_{\mathrm{BK}}$.

In the sections \ref{27}\hspace{0.5mm}-\ref{90}, we have studied the quantization of the ``squared'' Berry-Keating operator $H_{\mathrm{BK}}^2$ on compact quantum graphs. Proposition \ref{ab3} provides the complete classification of all self-adjoint realizations of $H_{\mathrm{BK}}^2$ again in terms of matrices $A$ and $B$ satisfying in this case the conditions \eref{47}. For the discrete spectrum of $H_{\mathrm{BK}}^2$, we have given in proposition \ref{ab4} the corresponding secular equation for $\lambda=k^2\neq0$. The zero eigenvalue $\lambda=k^2=0$ of $H_{\mathrm{BK}}^2$ plays a special r\^ole which we have discussed in section \ref{90} leading to the additional secular equation \eref{91}. Furthermore, in section \ref{ab8} we have discussed the conditions under which $H_{\mathrm{BK}}^2$ is the square of $H_{\mathrm{BK}}$ in the sense of \eref{36}, see proposition \ref{ab21}.

Based on the results derived in the previous sections, we have been able in section \ref{100} to state several theorems. In the theorems \ref{11104} and \ref{104} we have given an exact trace formula for $H_{\mathrm{BK}}$ respectively $H_{\mathrm{BK}}^2$ for a large class of test functions $h(k)$ belonging to the space $H_r$ defined in definition \ref{102}. The trace formulae establish a deep connection between the eigenvalue spectra of $H_{\mathrm{BK}}$ respectively $H_{\mathrm{BK}}^2$ and the length spectra of the periodic orbits of the corresponding classical dynamics. 

As an important consequence of the trace formulae, we have derived the Weyl asymptotics for $H_{\mathrm{BK}}$ (for $H_{\mathrm{BK}}^2$ we have used results in \cite{BE:2008}). The Weyl asymptotics of these operators have been given in the theorems \ref{m1} and \ref{110}, respectively. A comparison with the expected Weyl asymptotics \eref{505} respectively \eref{2} for the nontrivial Riemann zeros demonstrates clearly that neither $H_{\mathrm{BK}}$ nor $H_{\mathrm{BK}}^2$ can yield as eigenvalues the nontrivial zeros of the Riemann zeta function if these operators are self-adjoint realizations on any compact quantum graph, see theorem \ref{150}.

Finally, we have presented in section \ref{120} four simple examples illustrating some aspects of the quantum dynamics of $H_{\mathrm{BK}}$ respectively $H_{\mathrm{BK}}^2$.
\subsection*{Acknowledgements}
We would like to thank Wolfgang Arendt, Jens Bolte and Jan Eric Sträng for very helpful discussions and fruitful hints and Wolfgang Arendt for drawing our attention to \cite{Arendt:1994,Arendt:1995,Arendt:2002}. S.\ E.\ would like to thank the graduate school ``Analysis of evolution, information
and complexity'' of the Land Baden-Württemberg for the stipend which has enabled this paper.
\\
\appendix
{\small
\bibliographystyle{unsrt}
\bibliography{B.-K.-O.-head}}
\parindent0em
\end{document}